\documentclass[12pt]{iopart}
\usepackage{appendix}
\usepackage{graphicx}
\usepackage{color}
\usepackage[plainpages=false,pdfpagelabels,colorlinks=true,linkcolor=black,urlcolor=blue,citecolor=blue,pdftitle={Title},pdfauthor={},pdfdisplaydoctitle=true,pdfduplex=DuplexFlipLongEdge]{hyperref}

\begin{document}

\review[Short title]{Generalized Gibbs ensemble in integrable lattice models}

\author{Lev Vidmar and Marcos Rigol}

\address{Department of Physics, The Pennsylvania State University, University Park, PA 16802, USA}
\ead{luv2@psu.edu, mrigol@phys.psu.edu}
\vspace{10pt}

\begin{abstract}
The generalized Gibbs ensemble (GGE) was introduced ten years ago to describe observables in isolated integrable quantum systems after equilibration. Since then, the GGE has been demonstrated to be a powerful tool to predict the outcome of the relaxation dynamics of few-body observables in a variety of integrable models, a process we call generalized thermalization. This review discusses several fundamental aspects of the GGE and generalized thermalization in integrable systems. In particular, we focus on questions such as: which observables equilibrate to the GGE predictions and who should play the role of the bath; what conserved quantities can be used to construct the GGE; what are the differences between generalized thermalization in noninteracting systems and in interacting systems mappable to noninteracting ones; why is it that the GGE works when traditional ensembles of statistical mechanics fail. Despite a lot of interest in these questions in recent years, no definite answers have been given. We review results for the XX model and for the transverse field Ising model. For the latter model, we also report original results and show that the GGE describes spin-spin correlations over the entire system.
This makes apparent that there is no need to trace out a part of the system in real space for equilibration to occur and for the GGE to apply.
In the past, a spectral decomposition of the weights of various statistical ensembles revealed that generalized eigenstate thermalization occurs in the XX model (hard-core bosons).
Namely, eigenstates of the Hamiltonian with similar distributions of conserved quantities have similar expectation values of few-spin observables. Here we show that generalized eigenstate thermalization also occurs in the transverse field Ising model. 
\end{abstract}

%
%
%
\maketitle
\tableofcontents

\newpage

\section{Introduction}

Since the birth of statistical mechanics in the late 19th century, physicists have been fascinated by classical and quantum systems that cannot be understood using the traditional tools provided by this mighty framework. Of those extraordinary systems, our main focus in this review is on those that fall within the quantum realm. It is generally very difficult, usually impossible, to study the dynamics of macroscopic quantum systems far from equilibrium and to find their properties after they relax (if they do). The exact solution of two models of one-dimensional (1D) quantum magnets presented by Lieb, Schulz and Mattis in 1961~\cite{lieb61} (one of the models being the spin-1/2 XY spin model that we discuss later) provided the statistical physics community with powerful tools not only to study such systems in equilibrium, but also to explore their far-from-equilibrium dynamics. These models are prototypes of what we now know as integrable quantum systems \cite{sutherland_book_04}. After early studies of dynamics by Niemeijer~\cite{niemeijer67}, Mazur studied magnetization in the XY model and pointed out its nonergodic character~\cite{mazur69}. This was followed by a series of works by Barouch and coworkers~\cite{barouch69, barouch70, barouch71b}, which, using the modern terminology (to be introduced later), studied dynamics after quantum quenches. In those studies, it was apparent that traditional statistical mechanics did not describe observables after relaxation. It was later realized that conserved quantities in quantum spin chains play an important role in their transport properties, and that they may need to be taken into account for the interpretation of experiments in quasi-1D materials~\cite{castella95,zotos97}.
In particular, the possibility of ballistic heat (and eventually spin~\cite{prosen11}) transport at finite temperatures triggered many efforts to understand transport properties in 1D systems~(for reviews see, e.g., Refs.~\cite{zotos_prelovsek_04,heidrichmeisner07}, and also the review by Vasseur and Moore in this volume~\cite{Vasseur16}). 

A new burst of activity in 1D systems and their nonequilibrium dynamics has come with recent advances in cooling, trapping, and manipulating gases of atoms and molecules to reach the quantum degeneracy regime \cite{bloch08, cazalilla11}.
To constrain the dynamics to be effectively one dimensional, experimentalists use deep two-dimensional optical lattices \cite{moritz_stoferle_03, kinoshita04, paredes04} and atom chips \cite{folman00}.
Due to the high degree of isolation in those experiments (the gases are trapped by conservative potentials in ultrahigh vacuum), quantum coherence far from equilibrium can be preserved for long times, as demonstrated with the observation of collapse and revival phenomena in experiments with bosons \cite{greiner02, will_best_10} and fermions \cite{will15}. It is interesting to note that, despite the fact that ultracold quantum gases are not in contact with thermal reservoirs, it was for a long time taken for granted that their steady state could be described using traditional ensembles of statistical mechanics. A pioneering experiment by Kinoshita, Wenger, and Weiss~\cite{kinoshita06} demonstrated that this is not the case in 1D geometries. After taking an array of strongly interacting 1D Bose gases far from equilibrium, Kinoshita {\it et al}~showed that the steady state of the observables measured in the experiment was not the one expected in thermal equilibrium. Such steady states, now called prethermalized states, have also been obtained and thoroughly studied in a remarkable set of experiments with weakly interacting Bose gases in atom chips \cite{hofferberth07,gring12,langen15} (see also the review by Langen, Gasenzer, and Schmiedmayer in this volume~\cite{langen16}). A feature that these experimental setups have in common is their closeness to integrability. Other recent experiments that have studied near-integrable dynamics have dealt with strongly interacting bosons on 1D lattices~\cite{ronzheimer13,vidmar15}, as well as with Ising~\cite{simon11,meinert13} and Heisenberg~\cite{fukuhara13a,fukuhara13b,hild14} spin chains.

Shortly after the Kinoshita {\it et al}~experiments, it was shown that, in a system of hard-core bosons far from equilibrium, observables exhibit relaxation to time-independent values that can be predicted by means of a generalized Gibbs ensemble (GGE) \cite{rigol07,rigol06}. The GGE was obtained by maximizing the entropy, \`a la Jaynes \cite{jaynes57a,jaynes57b}, taking into account an extensive set of conserved quantities that made that particular hard-core boson model integrable.
Since then, the validity of the GGE to describe observables in integrable systems after equilibration has been demonstrated in a wide range of 1D models including:  Luttinger liquids \cite{cazalilla06,iucci_cazalilla_09,cazalilla12}, the $1/r$ fermionic Hubbard model \cite{kollar08}, the sine-Gordon model \cite{iucci_cazalilla_10}, the transverse field Ising model \cite{calabrese_essler_11, cazalilla12, calabrese_essler_12a, calabrese_essler_12b,  essler_evangelisti_12,fagotti13b, fagotti13, bucciantini14}, hard-core bosons in quasiperiodic lattices \cite{gramsch12}, hard-core anyons \cite{wright14}, bosons with contact interactions \cite{konik12, mossel12, collura13, collura13b, pozsgay_14b, sotiriadis14, Goldstein2015}, quantum field theories \cite{calabrese07, fioretto_mussardo_10, mussardo_13, cardy_16}, and spin-1/2 XXZ chains \cite{pozsgay_13, fagotti_collura_14, mierzejewski_prelovssek_14, wouters_denardis_14, pozsgay_mestyan14, ilievski15b}.

The works mentioned above have addressed several important questions related to the GGE. Among those, some that we touch upon in this review are: Why and which observables relax to the GGE predictions? Is the locality of an observable essential for it to be described by the GGE? Are there fundamental differences between noninteracting models and interacting models mappable to noninteracting ones? Which conserved quantities should one use to construct the GGE? Under which circumstances does the GGE description break down? No definite answers have been given to those questions. 

Here, we review numerical results and report original ones that provide answers to some of those questions in cases for which no analytic results are known.
More specifically, we focus on two paradigmatic integrable models that can be mapped onto noninteracting ones: hard-core bosons (XX model) and the transverse field Ising model (see also the review by Essler and Fagotti in this volume~\cite{essler16}). For the latter model, we compute the spin-spin correlations in the entire system both in the so-called diagonal ensemble and in the GGE, and show that the trace distance of such correlations in both ensembles vanishes in the thermodynamic limit. This is the second instance known to us in which correlation functions in an entire system are shown to be described by the GGE prediction (see Ref.~\cite{wright14} for the first one). This demonstrates that the prevailing view that relaxation to the GGE prediction occurs only in real-space subsystems of isolated quantum systems is not justified. Moreover, our study of the statistical weights of the eigenstates of the Hamiltonian reveals that, for every observable studied, nonvanishing weights are only present in a region around the system's mean energy in which eigenstates have similar expectation values of observables. These results support the generalized eigenstate thermalization scenario discussed in Ref.~\cite{cassidy11}, which can be thought of as a generalization of the eigenstate thermalization hypothesis~\cite{deutsch91, srednicki94, rigol08} to integrable systems. It provides a microscopic understanding for the general success of the GGE.

The presentation is organized as follows. In Sec.~\ref{models}, we introduce the models, the statistical ensembles, and specify the conserved quantities considered. Our exposition focuses on models that are mappable to noninteracting ones, which, as we argue in Sec.~\ref{relaxation}, should not be confused with noninteracting models. In Sec.~\ref{relaxation}, we present a theoretical discussion of the relaxation dynamics of experimentally relevant observables in a hard-core boson system. We show that values of the observables after relaxation are, up to finite size effects, those predicted by the GGE. In that section, we also introduce measures to quantify the differences between results after relaxation and the GGE predictions, and discuss their scaling with system size. Furthermore, we review the case of noninteracting fermions for which some observables fail to equilibrate but their time averages are still predicted by the GGE. Section~\ref{diagonal_gge} is devoted to the analysis of various ensembles after quenches in the transverse field Ising model. We report results for the diagonal ensemble, the GGE, and the grand canonical ensemble, and show how the results for observables in the GGE converge to those in the diagonal ensemble with increasing system size. Section~\ref{geth} is mainly devoted to a discussion of generalized eigenstate thermalization in the transverse field Ising model. We conclude with a summary and outlook in Sec.~\ref{conclusion}.

\section{Models, Ensembles, and Conserved Quantities} \label{models}

This section is devoted to the introduction of the models and statistical ensembles that are the focus of this review. We also show results for the distributions of conserved quantities after various quenches. Those quantities, and their expectation values in the initial state, are the core objects for the construction of the GGE.

\subsection{The XY model in the presence of a transverse magnetic field}

The XY model in the presence of a transverse magnetic field \cite{lieb61}, and, specially, some of its limits to be discussed below, are among the most studied models of quantum magnetism in 1D chains. The Hamiltonian can be written as:
\begin{equation}\label{eq:XYmodel}
\hat H_{\rm XY} = -J \sum_j \left[ (1+\gamma) \hat S_j^x \hat S_{j+1}^x + (1-\gamma) \hat S_j^y \hat S_{j+1}^y \right] - h  \sum_j\hat S_j^z 
\end{equation}
where $\hat S^x$, $\hat S^y$ and $\hat S^z$ are standard spin-1/2 operators, $J$ is the exchange constant, $h$ is the transverse magnetic field, and $\gamma$ is the anisotropy parameter. The Hamiltonian (\ref{eq:XYmodel}) is defined on a one-dimensional lattice with $L$ sites, with either open (in the first sum, $j=1,2,\ldots, L-1$) or periodic (in the first sum, $j=1,2,\ldots, L$, with $\hat S_{L+1}^\alpha\equiv \hat S_{1}^\alpha$) boundary conditions. We express the spin operators in terms of ladder operators $\hat S^{\pm}_j = \hat S^x_j \pm i\hat S^y_j$, which, in turn, can be expressed in terms of hard-core bosons as \cite{cazalilla11, holstein_primakoff_40}: $\hat S_j^+ = \hat b_j^\dagger \sqrt{1- \hat n_j}$, $\hat S_j^- =  \sqrt{1- \hat n_j}\, \hat b_j$, and $\hat S_j^z = \hat n_j - 1/2$, where $\hat n^{}_j = \hat b_j^\dagger \hat b^{}_j$. In the hard-core boson language, the Hamiltonian (\ref{eq:XYmodel}) can be written as 
\begin{equation} \label{Hxy_def}
\hat H_{\rm XY} = -\frac{J}{2} \sum_j \left(\hat b_j^\dagger \hat b^{}_{j+1} + {\rm H.c.} \right) - \frac{\gamma J}{2} \sum_j \left(\hat b_j^\dagger \hat b_{j+1}^\dagger + {\rm H.c.} \right) - h \sum_j \hat b_j^\dagger \hat b^{}_j+\frac{hL}{2},\
\end{equation}
where the hard-core boson operators satisfy a local constraint $(\hat b_j)^2 = (\hat b_j^\dagger)^2 = 0$. One can see that, in the hard-core boson language, $\gamma = 0$ is special. In that limit, the Hamiltonian commutes with the total number of bosons ($\hat N=\sum_j \hat b_j^\dagger \hat b^{}_j$), i.e., it is particle number conserving. This is known as the XX model in the spin language. In what follows, we treat the cases $\gamma = 0$ and $\gamma=1$ separately.

Before focusing on the different limits of the Hamiltonian (\ref{Hxy_def}), let us comment on whether we should consider this model as interacting or not. Hard-core bosons in Eq.~(\ref{Hxy_def}) are interacting because of the local constraint that precludes multiple occupancy of lattice sites. However, as we show in the next subsections, this model can be mapped onto a noninteracting one. Because of this mapping, the model in Eq.~(\ref{Hxy_def}) is sometimes referred to as a noninteracting model. An important point we want to stress in this review is that there are fundamental differences when it comes to equilibration and generalized thermalization between hard-core bosons (or spins) and the noninteracting fermions to which they can be mapped (as we know there are fundamental differences between their momentum distribution functions in equilibrium \cite{rigol05b}). For noninteracting fermions there can exist extensive sets of one-body observables that do not equilibrate, while this is not the case for hard-core bosons (in the absence of localization due to disorder) \cite{wright14}. How can such a fundamental difference be understood considering that there is a mapping between them? Mathematically, this follows from the fact that the mapping [see Eq.~(\ref{jordan_wigner})] is a nonlocal one. Physically, we can understand that such differences emerge because the one-body sector of a many-body system of noninteracting fermions evolves unitarily, while, because of interactions, this is not the case for the one-body sector of a many-body system of hard-core bosons. We will discuss this further in the context of one example in Sec.~\ref{relaxation}.

\subsubsection{The XX model: Hard-core bosons and noninteracting fermions.}
When $\gamma = 0$, particle number conservation implies that the last two terms in Eq.~(\ref{Hxy_def}) are constants that can be considered as an overall chemical potential. They are ignored in the discussion that follows for the XX model. One can generalize the hard-core boson Hamiltonian to account for a position dependent magnetic field in Eq.~(\ref{eq:XYmodel}), by adding a site-dependent potential $V_j$ (we define $\tilde J\equiv J/2$)
\begin{equation} \label{Hxx_def}
\hat H_{\rm XX} = - \tilde J  \sum_j \left(\hat b_j^\dagger \hat b^{}_{j+1} + {\rm H.c.}\right)  + \sum_j V_j \, \hat b_j^\dagger \hat b^{}_j.
\end{equation}
This is a model that can be realized by experimental groups using ultracold bosonic atoms in optical lattices in the limit of very strong onsite repulsive interactions between the bosons \cite{bloch08, cazalilla11}. It can be straightforwardly solved using the Jordan-Wigner transformation~\cite{cazalilla11,jordan_wigner_28}, which maps spin-1/2 operators (and therefore hard-core bosons) onto spinless fermions:
\begin{equation} \label{jordan_wigner}
\hat S^+_j = \hat f_j^\dagger e^{-i\pi \sum_{l<j} \hat f_l^\dagger \hat f^{}_l},\quad \hat S^-_j = e^{i\pi \sum_{l<j} \hat f_l^\dagger \hat f^{}_l} \hat f^{}_j, \quad \textrm{and}\quad \hat S_j^z = \hat f_j^\dagger \hat f^{}_j - 1/2,
\end{equation}
where $\hat f^{}_j$ and $\hat f_j^\dagger$ obey standard fermionic algebra. In the spinless fermion language, the Hamiltonian~(\ref{Hxx_def}) reads:
\begin{equation} \label{Hsf_def}
\hat H_{\rm XX} = - \tilde J  \sum_j \left(\hat f_j^\dagger \hat f^{}_{j+1} + {\rm H.c.}\right)  + \sum_j V_j \, \hat f_j^\dagger \hat f^{}_j.
\end{equation}

In the presence of translational invariance, $V_j=0$ (or constant) and periodic boundary conditions, a Fourier transform $\hat f^{}_j = 1/\sqrt{L} \sum_k e^{ikj} \hat f^{}_k$ diagonalizes the Hamiltonian~(\ref{Hsf_def}): $\hat H_{\rm XX} = -2\tilde J \sum_k \cos(k) \hat f_k^\dagger \hat f^{}_k$. The many-body eigenstates are hence products of single-particle states, $\prod_k\hat f_k^\dagger|\emptyset\rangle$, and the occupation of those single-particle states, $\hat m_k^f = \hat f_k^\dagger \hat f^{}_k$, are constants of motion. This can be straightforwardly extended to systems that are not translationally invariant by, instead of performing a Fourier transform, numerically diagonalizing the Hamiltonian~(\ref{Hsf_def}).

Due to its experimental relevance, the main observable that we consider in the context of $\hat H_{\rm XX}$ is the hard-core boson quasi-momentum distribution function
\begin{equation}
\hat m_k = \frac{1}{L} \sum_{j,l} e^{-i(l-j)k} \hat b_j^\dagger\hat b^{}_l.
\end{equation}
Its expectation value can be measured in experiments with ultracold quantum gases by means of the time-of-flight protocol~\cite{bloch08}. Since we only deal with lattice systems, in the reminder of this review we will refer to the quasi-momentum as the momentum. Because of the nonlocal character of the mapping between hard-core bosons and spinless fermions, and as mentioned before, $\langle\hat m_k\rangle$ is fundamentally different in those two systems \cite{rigol05b}. Once the fermionic problem has been solved, it is still challenging to obtain the momentum distribution of the hard-core bosons. Numerically, this can be achieved efficiently using properties of Slater determinants, as demonstrated for the ground state \cite{rigol04a,rigol_muramatsu_05}, finite temperature \cite{rigol05b}, and for the quantum dynamics \cite{rigol05a}. In contrast to the momentum distribution, the site occupations are identical for hard-core bosons and spinless fermions, $\langle\hat n_j\rangle = \langle\hat b_j^\dagger \hat b^{}_j\rangle = \langle\hat f_j^\dagger \hat f^{}_j\rangle$, and can therefore be calculated much more easily.

\subsubsection{The transverse field Ising model.}

The transverse field Ising model is obtained by substituting $\gamma = 1$ in Eqs.~(\ref{eq:XYmodel}) and (\ref{Hxy_def}). For this model, which is not particle number conserving in the hard-core boson language, we restrict our analysis to the translationally invariant case. Using the Jordan-Wigner and a Fourier transformation on the resulting fermionic Hamiltonian leads to:
\begin{eqnarray} \label{Hxy_sf}
\hat H_{\rm TFI}^{(+)}& =& \sum_{k \in {\cal K^{(+)}}} \left[ a_k \left( \hat f_k^\dagger \hat f^{}_k + \hat f_{-k}^\dagger \hat f^{}_{-k}  - 1\right) - b_k \left(i \hat f_k^\dagger \hat f_{-k}^\dagger + {\rm H.c.}\right)  \right],\nonumber \\ 
H_{\rm TFI}^{(-)}& =& \sum_{k \in {\cal K^{(-)}}>0} \left[ a_k \left( \hat f_k^\dagger \hat f^{}_k + \hat f_{-k}^\dagger \hat f^{}_{-k}  - 1\right) - b_k \left(i \hat f_k^\dagger \hat f_{-k}^\dagger + {\rm H.c.}\right)  \right]\\ &&- (J+h) \hat f_0^\dagger \hat f^{}_0 + (J-h) \hat f_\pi^\dagger \hat f^{}_\pi +h,\nonumber
\end{eqnarray}
where $a_k = -J\cos(k) - h$ and $b_k = J \sin(k)$ [for $\gamma\neq1$, $b_k = \gamma J \sin(k)$]. The sectors with even (+) and odd (-) number of fermions (from now on referred to as the even and odd sectors, respectively) are uncoupled and are treated separately. We note that, in the even sector, antiperiodic boundary conditions need to be used when mapping spins (hard-core bosons) onto spinless fermions \cite{lieb61}. As a result, the even and odd sectors are diagonalized in terms of different sets of wave vectors ${\cal K^{(+)}} = \{ \pi/L + n 2\pi/L \; \vert \; n=0,1,...,L/2-1 \}$ and ${\cal K^{(-)}}= \{ n 2\pi/L \; \vert \; n=0,...,L/2-1 \}$, respectively. In Eqs.~(\ref{Hxy_sf}), the wave vectors are coupled in pairs $\{ k,-k \}$. The only exception is found in the odd sector, in which a pair is formed by $k=0$ and $k=\pi$. As seen in Eqs.~(\ref{Hxy_sf}), this pair is treated separately. Formally, the Hamiltonian can be expressed as $\hat H_{\rm TFI} = \hat H_{\rm TFI}^{(+)} \hat {\cal P}^{(+)} + \hat H_{\rm TFI}^{(-)} \hat{\cal P}^{(-)}$, where the operators $\hat {\cal P}^{(\pm)}$ act as projectors onto a given sector~\cite{katsura62}. 

The Hamiltonians~(\ref{Hxy_sf}) are diagonalized by a Bogoliubov transformation $\hat f^{}_k = u_k \hat \eta^{}_k - v_k^* \hat \eta_{-k}^\dagger$, where $u_k = (\varepsilon_k + a_k)/\sqrt{2\varepsilon_k (\varepsilon_k + a_k)}$ and $v_k = i b_k/\sqrt{2\varepsilon_k (\varepsilon_k + a_k)}$. For $k=0$ and $k=\pi$, the original Hamiltonian is already diagonal, we take $\hat f_k = \hat \eta_{-k}^\dagger$. This results in:
\begin{eqnarray} \label{Hxy_diag}
\hat H_{\rm TFI}^{(+)}  = \sum_{k \in {\cal K}^{(+)}} \left[ \varepsilon_k (\hat \eta_k^\dagger \hat\eta^{}_k + \hat\eta_{-k}^\dagger \hat\eta^{}_{-k}) - \varepsilon_k \right],\\
\hat H_{\rm TFI}^{(-)}  = \sum_{k \in {\cal K}^{(-)}>0} \left[ \varepsilon_k (\hat \eta_k^\dagger \hat\eta^{}_k + \hat\eta_{-k}^\dagger \hat\eta^{}_{-k}) - \varepsilon_k \right]+ (h+J) \hat \eta_0^\dagger \hat \eta^{}_0 + (h-J) \hat \eta_\pi^\dagger \hat \eta^{}_\pi - h,\nonumber
\end{eqnarray}
with the energies of the noninteracting Bogoliubov quasiparticles being
\begin{equation} \label{esingle}
\varepsilon_k = \sqrt{h^2 + 2hJ \cos{k} + J^2}.
\end{equation}
[For $\gamma\neq1$, one has an extra term in the sum inside the square root in Eq.~(\ref{esingle}), which reads $(\gamma^2 - 1)J^2 \sin^2(k)$].

The transverse field Ising model has a quantum phase transition at $h=1$ between a ferromagnetic ground state ($h<1$) and a paramagnetic one ($h>1$)~\cite{sachdevbook}. The ferromagnetic ground state is doubly degenerate in the thermodynamic limit (each sector contributes one state). In finite systems, the ferromagnetic ground state is nondegenerate and belongs to the even sector. The paramagnetic ground state is nondegenerate (both, in finite systems and in the thermodynamic limit) and also belongs to the even sector. Whenever nondegenerate, the ground state $|0\rangle$ is the vacuum for Bogoliubov quasiparticles:
\begin{equation} \label{ngs}
|0\rangle = \prod_{k\in{\cal K}^{(+)}} \frac{1}{|v_k|} \hat \eta_k \hat \eta_{-k} |\emptyset\rangle \equiv | 0,0 \rangle \otimes \, ... \, \otimes | 0,0 \rangle  \otimes ...\ .
\end{equation}
All the eigenstates of the Hamiltonian (\ref{Hxy_diag}) can be obtained by acting with products (with the proper set of wave vectors) of $ \hat\eta_k^\dagger$ on the state $|0\rangle$. As a result, one can express any eigenstate of the Hamiltonian as
\begin{equation} \label{neigen}
|n\rangle = | p^{[n]}_{k_1}, p^{[n]}_{-k_1} \rangle \otimes \, ... \, \otimes | p^{[n]}_{k_j}, p^{[n]}_{-k_j} \rangle  \otimes ...\ .
\end{equation}
In the expressions above, $|p^{[n]}_{k}, p^{[n]}_{-k} \rangle$ denotes the occupation of Bogoliubov quasiparticles with $k$ and $-k$ in the $n$-th eigenstate of the Hamiltonian in a given sector. Each $\{k,-k\}$ subspace is spanned by the four vectors $\{ |0,0\rangle, |1,1\rangle, |1,0\rangle, |0,1\rangle \}$. The occupation of the Bogoliubov quasiparticles are constants of motion in the transverse field Ising model.

\subsection{The XXZ model}

For completeness, we also present the Hamiltonian of the XXZ model:
\begin{equation} \label{Hxxz_def}
\hat H_{\rm XXZ} = -J \sum_j \left(  \hat S_j^x \hat S_{j+1}^x + \hat S_j^y \hat S_{j+1}^y + \Delta \hat S_j^z \hat S_{j+1}^z \right),
\end{equation}
in which $\Delta$ sets the strength of the nearest-neighbor $\hat S^z\hat S^z$ interaction ($\Delta=1$ corresponds to the Heisenberg point). In contrast to XY model in a transverse field, the XXZ Hamiltonian~(\ref{Hxxz_def}) cannot be mapped onto a noninteracting model. Recent studies have shed light on new families of conserved quantities in this model~\cite{prosen11, ilievski15b, mierzejewski15, ilievski15a}
(see also the review by Ilievski, Medenjak, Prosen and Zadnik in this volume~\cite{ilievski16}).

\subsection{Quantum quenches and ensembles}

In the reminder of this review we will be interested in what happens to the systems introduced previously after they are taken far from equilibrium. A standard protocol used for the latter purpose is that of a quantum quench. Namely, the system is prepared in an eigenstate of a given time-independent Hamiltonian (usually, as done here, in the ground state), and the dynamics is studied under a new time-independent Hamiltonian (called $\hat{H}$ in what follows). This can be thought of as an instantaneous change in the parameters describing a system and, hence, the term quantum quench. 

Given the initial state $|\psi_0\rangle$, the time-evolving one can then be written as
\begin{equation}\label{eq:dynam1}
|\psi(t)\rangle  = e^{-i\hat Ht} |\psi_0\rangle = \sum_n e^{-iE_n t} |n \rangle\langle n| \psi_0\rangle = \sum_n e^{-iE_n t} c^{}_n | n \rangle,
\end{equation}
where $\{ | n \rangle \}$ is the complete set of eigenstates of $\hat H$ and $c^{}_n = \langle n| \psi_0\rangle$ is the projection of the initial state onto eigenstate $| n \rangle$ (we set $\hbar \equiv 1$). Rather than in the evolution of the wave-functions, here we are interested on how observables $\hat {\cal O}$ evolve under quantum dynamics
\begin{equation}\label{eq:dyn1}
\hspace*{-1.2cm} {\cal O} (t) \equiv \langle \psi(t) | \hat {\cal O} | \psi(t) \rangle = \sum_{n,m}^{E_n \neq E_m} e^{-i(E_n - E_m)t} c_m^* c^{}_n \langle m | \hat {\cal O} | n \rangle + \sum_{n, m}^{E_n = E_m} c_m^* c^{}_n \langle m | \hat {\cal O} | n \rangle.
\end{equation}
If an observable relaxes to a nearly time independent value (fluctuations about that value vanish with increasing system size and quantum revivals occur in time intervals that diverge with system size), i.e., if it equilibrates, then the equilibrated result for the observable is given (up to finite size effects) by the second term in the sum in Eq.~(\ref{eq:dyn1}). We note that this term includes both diagonal and off-diagonal matrix elements of degenerate eigenstates. 

For generic quantum systems one expects no degeneracies in the absence of special symmetries, so only diagonal matrix elements are expected to contribute to the results after relaxation, i.e., we expect ${\cal O}(t)$ to relax to $\langle\hat{\cal O}\rangle_{\rm DE}\equiv\textrm{Tr}[\hat\rho_{\rm DE}\hat{\cal O}]$, where
\begin{equation} \label{de_def}
\hat \rho_{\rm DE} = \sum_n \rho_{\rm DE}^{[n]}\, | n \rangle \langle n|
\end{equation}
is known as the diagonal ensemble density matrix \cite{rigol08}, and $\rho_{\rm DE}^{[n]}=|c_n|^2$. This has been shown to occur in numerical studies of nonintegrable systems \cite{rigol08, rigol_09a, rigol_09b, sorg14}. Degeneracies are expected to arise in integrable systems. However, in the thermodynamic limit, they usually do not lead to differences between the equilibrated results and the predictions of the diagonal ensemble \cite{cassidy11}. This, of course, unless the many-body spectrum of the integrable Hamiltonian has extensive degeneracies \cite{kollar08}. The latter is actually the case in the translationally invariant transverse field Ising model. In Sec.~\ref{subsec_Ik}, we will nevertheless show that, for the quenches studied, degenerate states play no role in the dynamics and the diagonal ensemble correctly predicts the equilibrated results for observables.

One of the central questions that has been studied in the last ten years is whether, after relaxation following a quantum quench, observables can be described using traditional ensembles of statistical mechanics. Since the results from the latter ensembles agree in the thermodynamic limit (at least for the observables that we are interested in), and since there is no particle-number conservation in the transverse field Ising model, the only traditional ensemble of statistical mechanics that we consider here (for both the XX and the transverse field Ising model) is the grand canonical ensemble. 

For the particle-number-conserving XX Hamiltonian~(\ref{Hxx_def}), we write
\begin{equation} \label{ge_xx}
\hat \rho_{\rm GE}^{\rm (XX)} = \frac{1}{Z_{\rm GE}^{\rm (XX)} } e^{-\beta(\hat H_{\rm XX}-\mu\hat N)},
\end{equation}
where $\hat N$ is the total particle number operator and $Z_{\rm GE}^{\rm (XX)}  = {\rm Tr} [ e^{-\beta (\hat H_{\rm XX} - \mu \hat N)} ]$. The values of $\beta$ and $\mu$ used to compare with the results from the quantum dynamics are set by constraining the mean energy and number of particles in the grand canonical ensemble to be the same as in the time-evolving state, i.e., $\langle \psi_0 | \hat H |\psi_0 \rangle = {\rm Tr} [ \hat \rho_{\rm GE}^{\rm (XX)}  \hat H]$ and $N = {\rm Tr} [ \hat \rho_{\rm GE}^{\rm (XX)}  \hat N]$.

For the (non-particle-number-conserving) transverse field Ising Hamiltonian~(\ref{Hxy_diag}), we have
\begin{equation} \label{ge_xy}
\hat \rho_{\rm GE}^{\rm (TFI)}  = \frac{1}{Z_{\rm GE}^{\rm (TFI)}} e^{-\beta \hat H_{\rm TFI}},
\end{equation}
where $Z_{\rm GE}^{\rm (TFI)} = {\rm Tr} [ e^{-\beta \hat H_{\rm TFI}} ]$ and $\beta$ is obtained by matching $\langle \psi_0 | \hat H |\psi_0 \rangle = {\rm Tr} [ \hat \rho_{\rm GE}^{\rm (TFI)}  \hat H]$. In what follows, expectation values of observables in the grand canonical ensemble are denoted as $\langle\hat{\cal O}\rangle_{\rm GE}\equiv\textrm{Tr}[\hat\rho_{\rm GE}\hat{\cal O}]$.

As already mentioned before, observables in integrable systems are not expected to relax to the predictions of the grand canonical ensemble. Instead, one expects them to relax to the predictions of the GGE, which is defined as
\begin{equation} \label{gge_def}
\hat \rho_{\rm GGE} = \frac{1}{Z_{\rm GGE}} e^{-\sum_k \lambda_k \hat I_k},
\end{equation}
where $\{ \hat I_k \}$ is a set of nontrivial conserved quantities that exists because the system is integrable. The corresponding partition function is $Z_{\rm GGE} = {\rm Tr}[e^{-\sum_k \lambda_k \hat I_k}]$. The Lagrange multipliers $\{ \lambda_k \}$ are fixed so that the GGE expectation value of each conserved quantity matches that in the initial state $\langle \psi_0 |\hat I_k | \psi_0 \rangle =  {\rm Tr} [ \hat \rho_{\rm GGE}  \hat I_k]$. We denote the expectation values of observables in the GGE as $\langle\hat{\cal O}\rangle_{\rm GGE}\equiv\textrm{Tr}[\hat\rho_{\rm GGE}\hat{\cal O}]$.

\subsection{Conserved quantities} \label{subsec_Ik}

\subsubsection{The XX model.}

\begin{figure}[!t]
\includegraphics[width=1\textwidth]{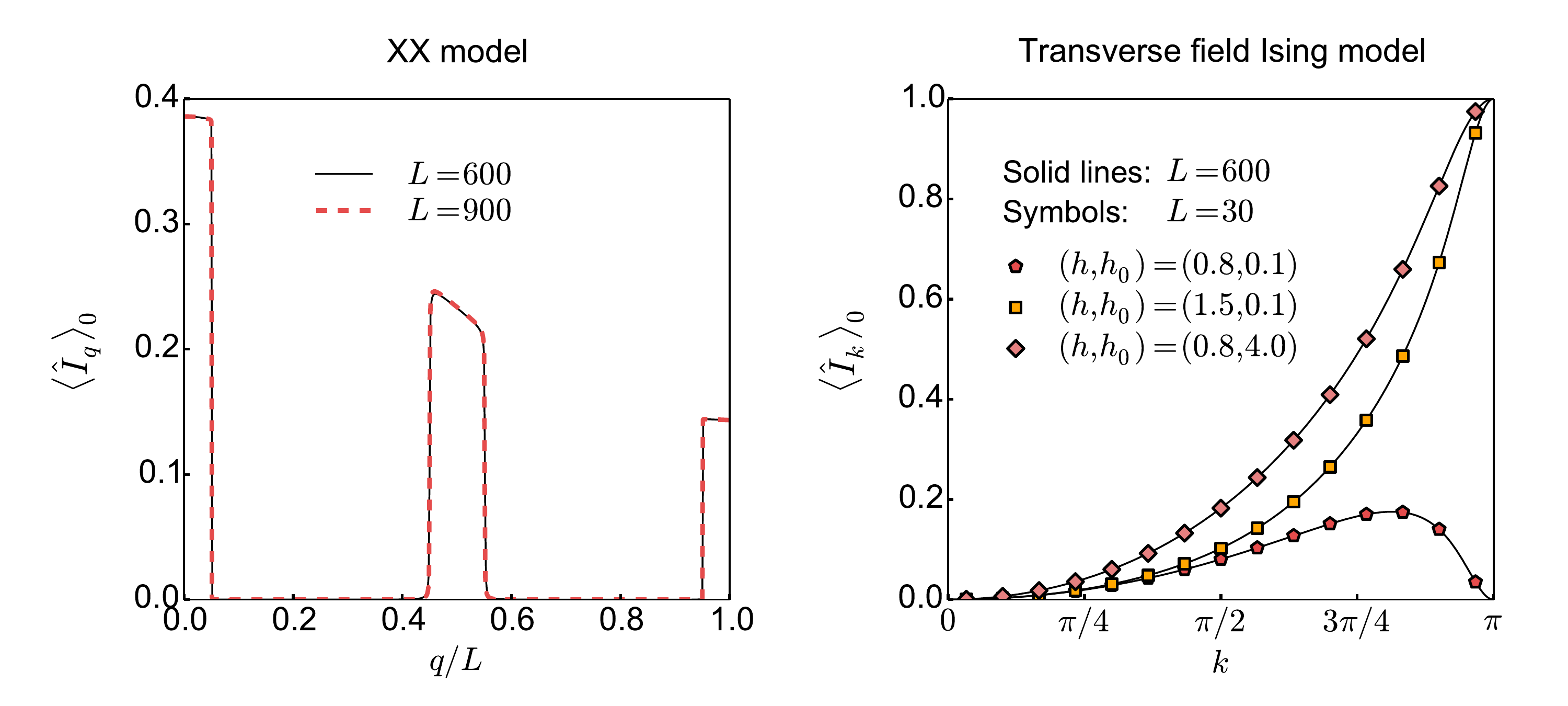}
\caption{
{\it Distribution of conserved quantities in the XX model and in the transverse field Ising model after a quantum quench.}
The conserved quantities $\langle \hat I_q \rangle_0$ in the XX model (left panel) are the occupations of the single-particle eigenstates of the fermionic Hamiltonian, where $q=1,2,\ldots,L$. The curves display $\langle \hat I_q \rangle_0$ after a quench from the ground state of $H_{\rm XX}$~(\ref{Hxx_def}) in the presence of a (superlattice) potential $V_j = A \cos(\frac{2\pi j}{T})$ with period $T=4$ and amplitude $A=8\tilde J$. The quench consists of turning off the superlattice potential. The average site occupancy is $N/L= 1/20$ \cite{rigol07}. The $q$-values are ordered with increasing eigenenergies of the single-particle eigenstates. We discuss further results for this quantum quench in Sec.~\ref{relaxation}. In the transverse field Ising model (right panel), the conserved quantities $\langle \hat I_k \rangle_0$ are the occupations of the Bogoliubov quasiparticles. The plots display $\langle \hat I_k \rangle_0$ after quenches from the ground state of a system with transverse field $h_0$ to a final field $h$. Results are not shown for $k<0$ since $\langle \hat I_{-k} \rangle_0 = \langle \hat I_k \rangle_0$. Further results for these quenches are discussed in Sec.~\ref{diagonal_gge}.
}
\label{fig_Ik}
\end{figure}

In the XX model, the set of conserved quantities $\hat I_q$ that we use to construct the GGE are the occupations of the single-particle eigenstates of the fermionic Hamiltonian (momentum modes occupations in a translationally invariant system). They can only take values zero or one, so the partition function of the GGE can be written as 
\begin{equation}
Z_{\rm GGE} = {\rm Tr}[  e^{-\sum_q \lambda_q \hat I_q}] = \prod_q (1+e^{-\lambda_q}),
\end{equation}
where $q=1,2,\ldots,L$. The expectation values of conserved quantities in the GGE can be written as
\begin{equation}
\langle \hat I_q  \rangle_{\rm{GGE}} = Z_{\rm GGE}^{-1} {\rm Tr} [ e^{-\sum_q \lambda_q \hat I_q}  \hat I_q] = \frac{e^{-\lambda_q}}{1+e^{-\lambda_q}}.
\end{equation}
Since the GGE is constructed requiring that those expectation values are the same as in the initial state $\langle \hat I_q  \rangle_0= \langle \psi_0 |\hat I_q | \psi_0 \rangle$, one obtains the following expression for the Lagrange multipliers \cite{rigol07}
\begin{equation} \label{Ikgge}
\lambda_q = \ln\left( \frac{1- \langle \hat I_q  \rangle_0}{\langle \hat I_q  \rangle_0} \right),
\end{equation}
and the partition function can then be written as
\begin{equation} \label{zgge}
Z_{\rm GGE}^{-1} = \prod_q (1 - \langle \hat I_q  \rangle_0).
\end{equation}
The left panel of Fig.~\ref{fig_Ik} displays the distribution of conserved quantities $\langle \hat I_q \rangle_0$ after turning off a superlattice potential in the XX model (with open boundary conditions)  \cite{rigol07}. The system was initially in the ground state in the presence of the superlattice potential. The dynamics and generalized thermalization after this quench will be discussed in Sec.~\ref{relaxation}. The number of values of $q$ is given by the number of lattice sites. They are ordered with increasing eigenenergies of the single-particle eigenstates, which are nondegenerate for open boundary conditions. As shown in the left panel in Fig.~\ref{fig_Ik}, a proper rescaling $q \to q/L$ makes the data for different system sizes collapse onto the same curve. 

\subsubsection{The transverse field Ising model. Initial eigenstate.}

In contrast to the XX model, for which the initial state in the expressions above can be any state (not necessarily an eigenstate of the Hamiltonian), for the transverse field Ising model we will restrict our analysis to initial states that are eigenstates of the Hamiltonian~(\ref{Hxy_diag}). In addition, we will focus on initial eigenstates that are in the even sector.

We are interested in the overlaps $c_n=\langle n|\psi_0 \rangle$ of the initial state $|\psi_0 \rangle$ with the eigenstates $\{|n\rangle\}$ of the final Hamiltonian. We write the latter using the notation in Eq.~(\ref{neigen}), and write the former as:
\begin{equation} \label{def_psi0}
|\psi_0\rangle = | r_{k_1}, r_{-k_1} \rangle \otimes \, ... \, \otimes | r_{k_j}, r_{-k_j} \rangle  \otimes ...\ .
\end{equation}
As a result, $c_n$ can be calculated as the product of the overlaps in each subspace $\{k,-k\}$, i.e., it can be written as a product of $L/2$ terms
\begin{equation}
c_n = \prod_{k \in {\cal K}^{(+)}} c_k^{[n]}.
\end{equation}
For each $k$, $c_k^{[n]}$ can take four possible values. This results from the fact that $|r_k,r_{-k}\rangle$ and $|p^{[n]}_k, p^{[n]}_{-k}\rangle$ are each one of $|0,0\rangle, |1,1\rangle, |1,0\rangle,$ and $|0,1\rangle$, which, in terms of spinless fermions, can be written as
\begin{equation}
\begin{array}{rcrcl}
|0,0\rangle & \equiv & \frac{1}{|v_k|} \hat \eta_k \hat \eta_{-k}|\emptyset\rangle & = & (-i) (u_k + v_k \hat c_k^\dagger \hat c_{-k}^\dagger) |\emptyset \rangle \\
|1,0\rangle & \equiv & \hat \eta_k^\dagger \left( \frac{1}{|v_k|} \hat \eta_k \hat \eta_{-k} \right) |\emptyset\rangle  & = &  (-i) \hat c_k^\dagger  |\emptyset \rangle \\
|0,1\rangle & \equiv & \hat \eta_{-k}^\dagger \left( \frac{1}{|v_k|} \hat \eta_k \hat \eta_{-k} \right) |\emptyset\rangle & = &  (-i) \hat c_{-k}^\dagger  |\emptyset \rangle \\
|1,1\rangle & \equiv & \hat \eta_k^\dagger \hat \eta_{-k}^\dagger \left( \frac{1}{|v_k|} \hat \eta_k \hat \eta_{-k} \right)|\emptyset\rangle  & = &i (v_k + u_k \hat c_k^\dagger \hat c_{-k}^\dagger ) |\emptyset \rangle
\end{array}.\label{eq:statsub}
\end{equation}
The dependence on the Hamiltonian parameters before and after the quench enters through the parameters $u_k=u_k(J,h)$ and $v_k=v_k(J,h)$, which were introduced with the Bogoliubov transformation [see Eq.~(\ref{Hxy_diag})]. 

Equation~(\ref{eq:statsub}) shows explicitly why an initial eigenstate that belongs to either the even or odd  sector has a nonzero overlap only with eigenstates in the same sector. On the level of $\{ k,-k \}$ subspaces, states $|0,0\rangle$ and $|1,1\rangle$ do not couple to the states $|1,0\rangle$ and $|0,1\rangle$, and vice versa. 

For the overlaps between the initial state and the eigenstates of the Hamiltonian we have that: \\
(i) If $r_{k}=0$ and $r_{-k}=0$, or $r_{k}=1$ and $r_{-k}=1$, then
\begin{equation}
c_k^{[n]} = \left\{ 
\begin{array}{lcll}
\pm \sqrt{\alpha_k} &\equiv& c_k^{(0)}& \mbox{if } \;  p^{[n]}_k = r_k, \;  \; p^{[n]}_{-k} = r_{-k}  \\ 
0                   &\equiv& c_k^{(1)}& \mbox{if } \;  p^{[n]}_k =1, \;  \; p^{[n]}_k =0  \\ 
0                   &\equiv& c_k^{(2)}& \mbox{if } \;  p^{[n]}_k =0, \;  \; p^{[n]}_{-k}=1  \\ 
\pm i \sqrt{1-\alpha_k} &\equiv& c_k^{(3)} & \mbox{if } \; p^{[n]}_k \neq r_k, \; \; p^{[n]}_{-k} \neq r_{-k} 
\end{array} \right. .\label{cnk1}
\end{equation}
(ii) If $r_{k}=1$ and $r_{-k}=0$, or $r_{k}=0$ and $r_{-k}=1$, then
\begin{equation}
c_k^{[n]} = \left\{ 
\begin{array}{lcll}
0 &\equiv& c_k^{(0)} & \hspace*{1.8cm} \mbox{if } \;  p^{[n]}_k=0, \;  \; p^{[n]}_{-k}=0\\ 
1 &\equiv& c_k^{(1)} & \hspace*{1.8cm} \mbox{if } \;  p^{[n]}_k = r_k, \;  \; p^{[n]}_{-k} = r_{-k}\\ 
0 &\equiv& c_k^{(2)} & \hspace*{1.8cm} \mbox{if } \;  p^{[n]}_k \neq r_k, \;  \; p^{[n]}_{-k} \neq r_{-k} \\
0 &\equiv& c_k^{(3)} & \hspace*{1.8cm} \mbox{if } \;  p^{[n]}_k=1, \;  \; p^{[n]}_{-k}=1
\end{array} \right. \label{cnk2} .
\end{equation}
Note that, in the expressions above, the superindex $[n]$ refers to the eigenstate number, while the superindex $(\xi)$, with $\xi=0,1,2,3$, labels one of the four possible values of $c_k^{[n]}$ for any given values of $r_{k}$ and $r_{-k}$ in the initial state. 

In Eq.~(\ref{cnk1}), we introduced the coefficient
\begin{equation} \label{def_alphak}
\alpha_k = \frac{1}{2} \left( 1+ \frac{a_k a_k^0 + b_k b_k^0}{\varepsilon_k \varepsilon_k^0} \right),
\end{equation}
which is a central quantity in the calculations that follow. The coefficients $a_k$ and $b_k$ were introduced in Eq.~(\ref{Hxy_sf}), and $\varepsilon_k$ is the single-particle energy defined in Eq.~(\ref{esingle}). We denote the parameters before the quench as $J_0$ and $h_0$ (they enter in $a_k^0$, $b_k^0$, and $\varepsilon_k^0)$, and the parameters after the quench as $J$ and $h$ (they enter in $a_k$, $b_k$, and $\varepsilon_k$). If not stated otherwise, we set $J=J_0=1$.

An important fact apparent from Eqs.~(\ref{cnk1}) and~(\ref{cnk2}) is the normalization of the weights $\sum_{\xi=0}^3\left| c_k^{(\xi)} \right|^2 = 1$, no matter the values of $r_{k}$ and $r_{-k}$ in the initial state. That this must be the case follows from the normalization of the initial state
\begin{equation} \label{cnsumrule}
\langle\psi_0|\psi_0\rangle=\sum_n |c_n|^2 = \sum_n \prod_{k \in {\cal K}^{(+)}} \left| c_k^{[n]} \right|^2 = \prod_{k \in {\cal K}^{(+)}} \left( \sum_{\xi=0}^3 \left| c_k^{(\xi)} \right|^2 \right) = 1.
\end{equation}

The conserved quantities we use to construct the GGE are the occupations of the Bogoliubov quasiparticles $\{ \hat I_k\equiv \hat\eta^\dagger_{k} \hat\eta^{}_{k}\}$. In the eigenstates of the Hamiltonian, $p_k^{[n]} = \langle n | \hat I_k | n \rangle$ can only take values 0 or 1. Using Eq.~(\ref{Ikgge}), this implies that
\begin{equation} \label{elambda}
\langle n|e^{-\lambda_k \hat I_k} |n\rangle = \left\{
\begin{array}{ll}
1 & {\rm if} \; p_k^{[n]}  = 0 \\ \frac{\langle \hat I_k \rangle_0}{1- \langle \hat I_k \rangle_0} & {\rm if} \; p_k^{[n]} =1
\end{array}
\right. .
\end{equation}

As for traditional ensembles of statistical mechanics, one can write the density matrix of the GGE as a sum over the contribution from all the eigenstates of the Hamiltonian, $\hat \rho_{\rm GGE} = \sum_n \rho_{\rm GGE}^{[n]} |n\rangle \langle n|$. Using Eqs.~(\ref{zgge}) and~(\ref{elambda}), the weights $\rho_{\rm GGE}^{[n]}$ can be expressed solely through the expectation values of the conserved quantities in the initial state
\begin{equation}
\rho_{\rm GGE}^{[n]} = \prod_{k \in\{ {\cal K}^{(+)},-{\cal K}^{(+)}\}}^{p_k^{[n]} = 0} \left(1 - \langle \hat I_k \rangle_0\right)  \prod_{k \in \{{\cal K}^{(+)},-{\cal K}^{(+)}\} }^{p_k^{[n]} = 1} \langle \hat I_k \rangle_0.
\end{equation}
Here, the index $k$ of the momentum runs through all the possibles values (positive and negative) in the lattice, which we denote as $k \in\{ {\cal K}^{(+)},-{\cal K}^{(+)}\}$.
 
Hence, in order to determine $\rho_{\rm GGE}^{[n]}$, we only need to compute the expectation values $\langle \hat I_k \rangle_0$. This can be achieved using the overlaps
\begin{equation} \label{Ikcn}
\langle \hat I_k \rangle_0 = \sum_{n,m} \langle \psi_0 | m \rangle \langle m |\hat  I_k | n \rangle \langle n | \psi_0 \rangle = \sum_n^{ p_k^{[n]} = 1} |c_n|^2,
\end{equation}
which means that the initial and final Hamiltonian parameters enter the GGE only through one parameter, namely, $\alpha_k$. Equation~(\ref{Ikcn}) can be further simplified using the normalization of single-particle weights [Eq.~(\ref{cnsumrule})]. One gets that $\langle \hat I_k \rangle_0$ is nothing but $\left| c_k^{(\xi)} \right|^2$, where $\xi$ is determined by $\{r_k,r_{-k}\}$. This leads to the general expression for the expectation values of conserved quantities:
\begin{equation} \label{Iksingle}
\langle\hat  I_k \rangle_0 = \left\{ 
\begin{array}{lcccc}
1-\alpha_k  & \mbox{if } & r_k = 0 & {\rm and} & r_{-k}  = 0 \\
\alpha_k & \mbox{if } &  r_k =1 & {\rm and} & r_{-k}  = 1 \\
1 & \mbox{if }  & r_k = 1  & {\rm and} & r_{-k} = 0 \\
0   & \mbox{if } & r_k  = 0 & {\rm and} & r_{-k} = 1
\end{array} \right. .
\end{equation}

An important clarification is in order at this point on the use of the GGE for finite XX and transverse field Ising systems. All expressions obtained so far for the GGE follow after a grand canonical trace. Such a trace is problematic in systems with periodic boundary conditions because the mapping between spins (hard-core bosons) and fermions requires that sectors with even number of fermions be treated with antiperiodic boundary conditions \cite{lieb61}. Instead, our expressions are obtained assuming that the boundary conditions are the same for sectors with even and odd particle numbers. To circumvent this problem, in all our calculations for the XX model we report results for systems with open boundary conditions, for which no such problem exists \cite{rigol05b}. For the  transverse field Ising model, for which periodic boundary conditions are chosen, the assumption that boundary conditions are the same for even and odd number of fermions (we use the one for even number of fermions) introduces an error. The effect of that error on the observables studied here decreases with increasing system size and vanishes in the thermodynamic limit, in which the boundary conditions become irrelevant, i.e., the error we have introduced in our calculations can be thought of as an additional finite-size effect.

\subsubsection{The transverse field Ising model. Initial ground state.}

In what follows we focus on quenches in finite systems from the ground state [$r_k = r_{-k} = 0$ in Eq.~(\ref{def_psi0})] for a field $h_0$ to a final field $h$. Hence, the initial state in all our quenches belongs to the even sector. For these quenches, only $\sqrt{\cal D}$ of the weights in the diagonal ensemble are nonzero, where ${\cal D} = 2^L$ is the total size of the Hilbert space. This is because any eigenstate of the final Hamiltonian with $... \, \otimes|1, 0\rangle\otimes...$ or $...\otimes|0,1\rangle\otimes...$ will have a vanishing overlap with the initial state. The nonzero weights are given by the expression
\begin{equation}\label{eq:overde}
\rho_{\rm DE}^{[n]} = \prod_{k\in {\cal K}^{+}}^{p_k^{[n]}=0} \alpha_k \prod_{k\in {\cal K}^{+}}^{p_k^{[n]}=1} (1-\alpha_k),
\end{equation}
which is a product of $L/2$ terms.
The eigenstate spectrum of the transverse field Ising model has extensive
degeneracies, which predominantly occur due to single occupancies of the $\{k ,-k\}$ subspaces (the two singly occupied states are degenerate). After the quench from an initial ground state, none of the $\{k ,-k\}$ subspaces is singly occupied. As a result, the eigenstates with nonzero weights in the diagonal ensemble are nondegenerate, except for some accidental (nonextensive) degeneracies.

The occupation of the conserved quantities~(\ref{Iksingle}) for an initial ground state simplifies to
\begin{equation} \label{Ik_xy_gs}
\langle \hat I_k \rangle_0 = (1-\alpha_k).
\end{equation}
The distribution of $\langle \hat I_k \rangle_0$ is shown in the right panel of Fig.~\ref{fig_Ik} following three quenches of the transverse field $h_0 \to h$. When quenching from the paramagnetic ($h_0>J$) to the ferromagnetic ($h<J$) side of the transverse field Ising model (or vice versa), the distribution increases monotonically with increasing $k$. In contrast, for quenches within the same phase, the distribution is peaked at some finite $k$ and then decreases towards zero for $k \to \pi$. The behavior as $k \to \pi$ can be understood from Eq.~(\ref{Ik_xy_gs}), which gives $\langle \hat I_{k\to\pi} \rangle_0 = [1-{\rm sgn}(h-J) {\rm sgn}(h_0-J)]/2$.

Equation~(\ref{Ik_xy_gs}) also simplifies the expression for the weights in the GGE
\begin{equation}\label{eq:overGGE}
\rho_{\rm GGE}^{[n]}
= \prod_{k \in {\cal K}^{(+)} }^{p_k^{[n]} = p_{-k}^{[n]} = 0} \alpha_k^2
\prod_{k \in {\cal K}^{(+)} }^{p_k^{[n]} \neq p_{-k}^{[n]}} \alpha_k (1- \alpha_k)
\prod_{k \in {\cal K}^{(+)} }^{p_k^{[n]} = p_{-k}^{[n]} = 1} (1- \alpha_k)^2.
\end{equation}

It is important to realize that, when quenching from the ground state, all eigenstates of the final Hamiltonian ($\cal D$ states) can potentially have nonzero weights in the GGE. In contrast, as mentioned before, only $\sqrt{\cal D}$ eigenstates of the Hamiltonian have nonzero weights in the diagonal ensemble. It is therefore not at all obvious that the GGE should describe observables after equilibration. An interesting observation can be made for eigenstates that have nonzero weights in both ensembles. While their weight in the DE is $|c_n|^2$, their weight in the GGE is $|c_n|^4$.

\subsubsection{Statistical independence of macroscopic subsystems in the GGE.}

We note at this point that, in $\hat\rho_{\rm GGE}$, the conserved quantities appear in the form $e^{-\lambda_k \hat I_k}$, as in traditional ensembles of statistical mechanics. However, the conserved quantities selected by us (occupations of single-particle states) are not extensive as in traditional ensembles of statistical mechanics. Extensivity is required for the factorizability of the density matrix, which is needed to ensure statistical independence of macroscopic subsystems. Instead, in the exponential in $\hat\rho_{\rm GGE}$, we have an extensive number of intensive conserved quantities. This might lead one to conclude that our GGE is not a legitimate statistical ensemble. 

There are two ways to see that our GGE leads to statistical independence of macroscopic subsystems. As shown in Fig.~\ref{fig_Ik}, the expectation values of the conserved quantities after the quenches are smooth functions (independent of the system size) of $q/L$ for the XX model and $k$ for the transverse field Ising model. Because of Eq.~(\ref{Ikgge}), the Lagrange multipliers are also smooth functions of $q/L$ and $k$, respectively. One can then define extensive (coarse grained) integrals of motion $\hat {\cal I}_{\alpha}\equiv\sum_{\alpha' \in [\alpha-\delta_\alpha/2, \alpha+\delta_\alpha/2 ]} \hat I_{\alpha'}$, where $\alpha\equiv q/L$ ($\delta_\alpha\ll1$) for the XX model and $\alpha\equiv k$ ($\delta_\alpha\ll\pi$) for the transverse field Ising model, such that $\hat \rho_{\rm GGE} \propto e^{-\sum_\alpha \lambda_\alpha \hat {\cal I}_{\alpha}}$ \cite{cassidy11,rigol_fitzpatrick_11}. Alternatively, extensive conserved quantities can be constructed by a linear transformation of $\{ \hat I_{k} \}$, which enables ordering them according to their support on the lattice~\cite{fagotti13}. This latter construction is particularly appealing because it allows one to show that the larger the support of an extensive conserved quantity is the least it affects expectation values of local observables in the GGE. If one wants to compute a local observable with a given finite accuracy, one can then truncate the GGE to have a finite number of local extensive conserved quantities~\cite{fagotti13}.

\subsection{Expectation values of observables in the transverse field Ising model} \label{sec:observables}

Let us discuss how to compute expectation values of observables in the transverse field Ising model in the various ensembles of interest. For that, we write the expectation value of an observable $\hat {\cal O}$ in an ensemble $\mu$ ($\mu=$DE, GE, or GGE), with weights \{$\rho_\mu^{[n]}$\}, as
\begin{equation} \label{Oformal}
\langle \hat {\cal O} \rangle_\mu = \sum_n \rho_{\mu}^{[n]} \langle n | \hat{\cal O} | n\rangle,
\end{equation}
This expression can be further simplified if two conditions are met: (i) the weights $\rho_\mu^{[n]}$ factorize (as they do in the ensembles considered in this review), $\rho_\mu^{[n]} = \prod_{k\in {\cal K}^{(+)}} \rho_{k,\mu}^{[n]}$, and (ii) the eigenstate expectation values can be written as sums of single-particle contributions, $\langle n | \hat{\cal O} | n\rangle = \sum_{k\in{\cal K}^{(+)}}\langle n | \hat{\cal O}_k | n\rangle$.  A single set of wave vectors needs to be chosen for the even and odd sectors. This, as we mentioned before, introduces an error whose effect in our observables of interest vanishes in the thermodynamic limit.

If the conditions above are met, we write
\begin{equation}\label{eq:Oxi}
 \langle n | \hat{\cal O}_k | n\rangle \equiv  \left\{ 
\begin{array}{lclcl}
{\cal O}_k^{(0)} & \mbox{if } & p_k^{[n]} = 0 & {\rm and} & p_{-k}^{[n]}  = 0 \\
{\cal O}_k^{(1)} & \mbox{if } & p_k^{[n]} = 1 & {\rm and} & p_{-k}^{[n]}  = 0 \\
{\cal O}_k^{(2)} & \mbox{if } & p_k^{[n]} = 0 & {\rm and} & p_{-k}^{[n]}  = 1 \\
{\cal O}_k^{(3)} & \mbox{if } & p_k^{[n]} = 1 & {\rm and} & p_{-k}^{[n]}  = 1 
\end{array} \right. .
\end{equation}
and
\begin{equation} \label{eqrhogs}
\rho_{k,\mu}^{[n]}  \equiv \left\{ 
\begin{array}{lclcl}
\rho_{k,\mu}^{(0)} & \mbox{if } & p_k^{[n]} = 0 & {\rm and} & p_{-k}^{[n]}  = 0 \\
\rho_{k,\mu}^{(1)} & \mbox{if } & p_k^{[n]} = 1 & {\rm and} & p_{-k}^{[n]}  = 0 \\
\rho_{k,\mu}^{(2)} & \mbox{if } & p_k^{[n]} = 0 & {\rm and} & p_{-k}^{[n]}  = 1 \\
\rho_{k,\mu}^{(3)} & \mbox{if } & p_k^{[n]} = 1 & {\rm and} & p_{-k}^{[n]}  = 1 
\end{array} \right. .
\end{equation}\label{cnk}
Note that the ordering of ${\cal O}_k^{(\xi)}$ and $\rho_{k,\mu}^{(\xi)}$ according to $\xi$ in the above equations follows the ordering in Eq.~(\ref{cnk1}) for the case of the initial ground state, $r_k = r_{-k}=0$.
Equations~(\ref{eq:Oxi}) and~(\ref{eqrhogs}), and the fact that the weights $\rho_{k,\mu}^{(\xi)}$ are normalized in the $\{k,-k\}$ subspaces, allows us to rewrite Eq.~(\ref{Oformal}) as
\begin{equation} \label{Osingle}
\langle \hat {\cal O} \rangle_\mu = \sum_{k \in {\cal K}^{(+)}} \left( \sum_{\xi = 0}^{3} \rho_{k,\mu}^{(\xi)} {\cal O}_{k}^{(\xi)} \right).
\end{equation}
We have already shown that the weights $\rho_{k,{\rm DE}}^{(\xi)}$ are normalized in the $\{k,-k\}$ subspaces. It is straightforward to show it for the GGE and the grand canonical ensemble. For example, for the GGE one can see from Eq.~(\ref{eq:overGGE}) that
\begin{equation}
\sum_{\xi = 0}^{3} \rho_{k,{\rm GGE}}^{(\xi)} =\alpha_k^2 + 2\alpha_k (1-\alpha_k) + (1-\alpha_k)^2 = 1.
\end{equation}

There is another important class of observables in the transverse field Ising model that can be straightforwardly evaluated in terms of single-particle contributions, but not as in Eq.~(\ref{Osingle}). These observables, we call them $\hat {\cal A}$, can be written as
\begin{equation} \label{O2formal}
\langle \hat {\cal A} \rangle_\mu = \sum_n \rho_\mu^{[n]} \langle n | \hat {\cal A}_1 | n \rangle  \langle n | \hat {\cal A}_2 | n \rangle,
\end{equation}
where the expectation values $\langle\hat {\cal A}_1 \rangle_\mu$ and $\langle \hat {\cal A}_2 \rangle_\mu$ can be computed using Eq.~(\ref{Osingle}). As a result, $\langle \hat {\cal A} \rangle_\mu$ reads:
\begin{equation} \label{O2single}
\hspace*{-2.5cm}
\langle \hat {\cal A} \rangle_\mu = \langle\hat {\cal A}_1 \rangle_\mu \langle\hat {\cal A}_2 \rangle_\mu
- \sum_{k \in {\cal K}^{(+)}} \left( \sum_{\xi = 0}^3 \rho_{k,\mu}^{(\xi)} {\cal A}_{1,k}^{(\xi)} \right) \left( \sum_{\xi' = 0}^3 \rho_{k,\mu}^{(\xi')} {\cal A}_{2,k}^{(\xi')} \right)
+ \sum_{k \in {\cal K}^{(+)}} \left( \sum_{\xi = 0}^3 \rho_{k,\mu}^{(\xi)} {\cal A}_{1,k}^{(\xi)} {\cal A}_{2,k}^{(\xi)} \right),
\end{equation}
where ${\cal A}_{1,k}^{(\xi)}$ and ${\cal A}_{2,k}^{(\xi)}$ are defined as ${\cal O}_k^{(\xi)}$ is in Eq.~(\ref{eq:Oxi}).

Equations~(\ref{Osingle}) and~(\ref{O2single}) allow us to calculate numerically, in polynomial time, various observables of interest for the transverse field Ising model, see Sec.~\ref{diagonal_gge}.

\section{Dynamics and the GGE in the XX model} \label{relaxation}

In this section, we discuss the dynamics in the XX model after a quantum quench. We focus mostly on hard-core bosons and compare the results after relaxation to the predictions of the grand canonical ensemble and the GGE. As mentioned in the introduction, it was on hard-core boson systems for which it was first established the relevance of the GGE to describe equilibrated values of observables in isolated integrable systems~\cite{rigol07,rigol06}. Here, we also discuss the behavior of noninteracting spinless fermions. We point out a fundamental difference between them and hard-core bosons in terms of equilibration and generalized thermalization.

\subsection{Dynamics and generalized thermalization of hard-core boson systems} \label{sec_hcb}

Some results for the first quench considered here have been already reported in Ref.~\cite{rigol07}. This quench was motivated by the experimental protocol followed by Kinoshita {\it et al} \cite{kinoshita06}. They pulsed a 1D lattice along an array of 1D Bose gases to take them far from equilibrium. The pulses generated peaks in the momentum distribution function of the bosons at multiples of the lattice wave-number. The oscillatory dynamics of those peaks was then studied by means of a special kind of time-of-flight measurements (the expansion was carried out in one dimension). Because of the observed oscillatory behavior of the peaks, this experiment is known as the quantum Newton's cradle.

\begin{figure}[!t]
\includegraphics[width=1\textwidth]{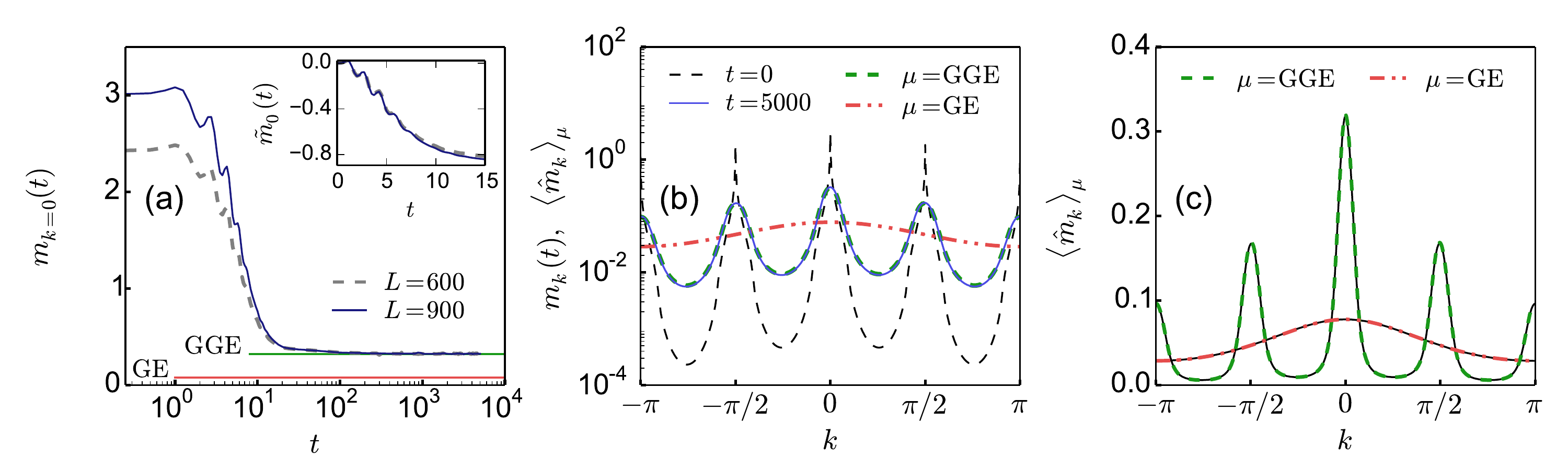}
\caption{
{\it Dynamics of the momentum distribution of hard-core bosons after a quantum quench and its description after relaxation.} The initial state is the ground state of $\hat H_{\rm XX}$, see Eq.~(\ref{Hxx_def}), with a superlattice potential $V_j = 8 \tilde J \cos(\pi j/2)$ and open boundary conditions. We consider systems with $N=30$ and 45 hard-core bosons in $L=600$ and 900 lattice sites, respectively, for which the average site occupancy is $n= N/L= 1/20$. At $t=0$, the superlattice potential is turned off and the system is let evolve unitarily under $\hat H_{\rm XX}$. (a) Time evolution of $m_{k=0}(t)$ for $L=600$ and 900. Horizontal lines denote the corresponding $\langle \hat m_{k=0}\rangle_{\rm GGE}$ and $\langle \hat m_{k=0}\rangle_{\rm GE}$ for $L=900$. The inset displays the rescaled short-time evolution of $\tilde{m}_0(t)$ (see text) as a function of $t$. (b) The entire momentum distribution in the initial state ($t=0$), at a long time after the quench ($t=5000$ in units of $\hbar/\tilde{J}$), as well as $\langle \hat m_k\rangle_{\rm GGE}$ and $\langle \hat m_k\rangle_{\rm GE}$, for $L=900$. (c) Results for $\langle \hat m_k\rangle_{\rm GGE}$ and $\langle \hat m_k\rangle_{\rm GE}$ in the system with $L=600$ (thin solid lines) and $L=900$ (thick dashed lines). The results for $L=600$ were taken from Ref.~\cite{rigol07}.
}
\label{fig_nk2007}
\end{figure}

The specific quench protocol we consider can be thought of as a simplified version of the experimental protocol. Our initial state is taken to be the ground state of hard-core bosons in the presence of a (superlattice) potential $V_j = A \cos(\frac{2\pi j}{T})$ [see Eq.~(\ref{Hxx_def})]. The period $T$ is taken to be $T=4$, and gives rise to four sharp peaks in the momentum distribution $m_k(t) = \langle\psi(t) | \hat m_k |\psi(t) \rangle$ at $t=0$, as shown in Fig.~\ref{fig_nk2007}(b). After the sudden turn off of the superlattice potential, $m_k(t)$ undergoes dynamics and relaxes to a new steady state distribution. We consider systems with 30 and 45 hard-core bosons in lattices with $L=600$ and 900 sites, respectively, i.e., the average site occupation is very low (0.05). Because of this (the average interparticle distance is much larger than the lattice spacing), one can think of these systems as being in the continuum, which was the case studied in the experiments. The conserved quantities $\langle \hat I_q \rangle_0$ we use to construct the GGE~[see Eq.~(\ref{gge_def})] are the occupations of the single-particle eigenstates of the final fermionic Hamiltonian (with open boundary conditions). Their distribution is shown in the left panel of Fig.~\ref{fig_Ik}.

In Fig.~\ref{fig_nk2007}(a), we show the time evolution of $m_{k=0}(t)$ for the two system sizes considered. Remarkably, both can be seen to relax to the same steady-state result despite the fact that, because of quasi-long range order \cite{rigol05a}, the initial value of $m_{k=0}$ is different in both systems. By plotting $\tilde{m}_0(t)=[m_{k=0}(t)-m_{k=0}(t=0)]/m_{k=0}(t=0)$ vs $t$, one can see [inset in Fig.~\ref{fig_nk2007}(a)] that the short time dynamics is essentially the same in both systems. Figure~\ref{fig_nk2007}(a) also shows that the result after relaxation can be predicted by the GGE, and that it is clearly different from the grand canonical ensemble prediction. As mentioned before, whenever observables relax to the GGE predictions [as in Fig.~\ref{fig_nk2007}(a)] we say that they exhibit generalized thermalization. We use the word ``generalized'' to differentiate this process from thermalization, namely, the relaxation to the predictions of traditional ensembles of statistical mechanics.

That generalized thermalization occurs in these systems is better seen in Fig.~\ref{fig_nk2007}(b), in which we plot the entire momentum distribution functions in the initial state, at $t=5000$ (in units of $\hbar/\tilde{J}$), as well as in the GGE and in the grand canonical ensemble. It is most remarkable that, after relaxation, the momentum distribution function of the hard-core bosons retains information about the initial state by exhibiting four clearly resolved peaks at the same positions as in the initial state. The GGE accurately describes the entire momentum distribution after relaxation. This is in stark contrast to the prediction from the grand canonical ensemble [see Eq.~(\ref{ge_xx})], which only exhibits a single broad peak (as expected in a system with no additional lattice). By comparing the results for systems with $L=600$ and $900$ lattices sites in Fig.~\ref{fig_nk2007}(c), one can see that finite-size effects are negligible for the system sizes considered, i.e., one could think of these results as being in the thermodynamic limit.

An important question that needs to be addressed at this point is how the time fluctuations of observables, about the steady state result, behave with increasing system size. For hard-core bosons, this was studied in Refs.~\cite{gramsch12, wright14, cassidy11}, while for noninteracting spinless fermions this was studied in Refs.~\cite{camposvenuti13, ziraldo_silva_12, ziraldo_santoro_13, he13}. In the absence of localization due to disorder or quasi-periodic potentials, a remarkable finding has been that all one-body observables studied for hard-core bosons exhibit time fluctuations that decrease with increasing system size (as $1/L^\kappa$, with $\kappa=1/2$ or 1). In contrast, some one-body observables for noninteracting spinless fermions do not equilibrate, while others exhibit time fluctuations that decrease with increasing system size (as $1/\sqrt{L}$). A decay of the time fluctuations as $1/\sqrt{L}$ is the characteristic scaling of the Gaussian equilibration scenario proposed in Ref.~\cite{camposvenuti13} for noninteracting spinless fermions.

\begin{figure}[!t]
\includegraphics[width=1\textwidth]{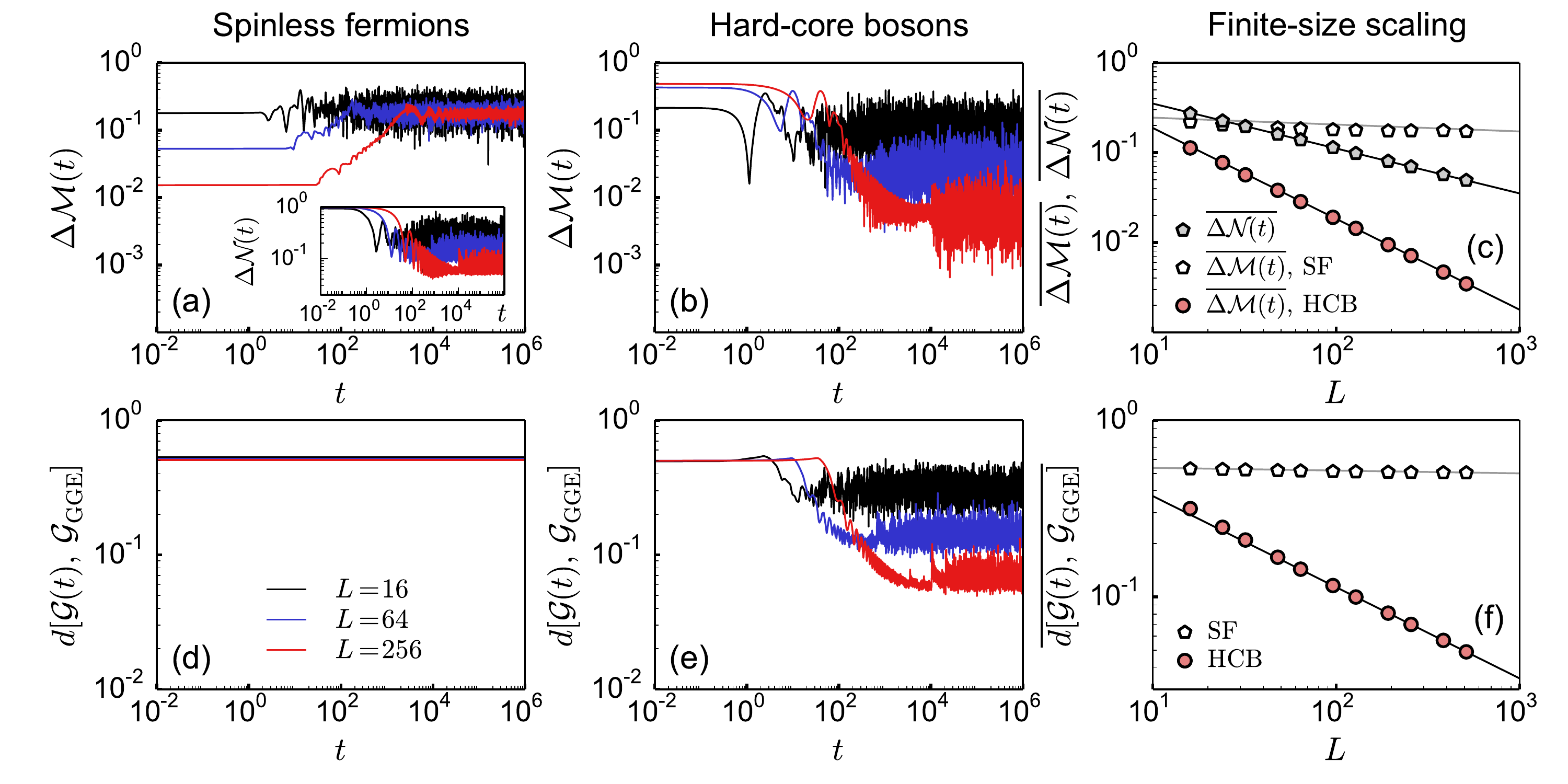}
\caption{
{\it Dynamics of the momentum distribution, site occupations and the trace distance after a quantum quench.}
The initial state is the ground state of $\hat H_{\rm XX}$, see Eq.~(\ref{Hxx_def}), in a box potential for a system with twice as many sites as particles $N$. The box is then opened in a larger box of size $L=4N$. (a),(b) Distance between the instantaneous and the GGE momentum distribution $\Delta {\cal M}(t)$, Eq.~(\ref{deltaM_def}), for spinless fermions (SF) and hard-core bosons (HCB), respectively. The inset in (a) shows the distance between the instantaneous and the GGE site occupations $\Delta {\cal N}(t)$, Eq.~(\ref{deltaN_def}). (c) The average values of $\Delta {\cal M}(t)$ and $\Delta {\cal N}(t)$, in the time interval $t \in [10^5,10^6]$, are plotted as a function of the system size $L$. Black solid lines are power-law fits $a L^{-\kappa}$ for $L \geq 96$, yielding $a=1.93$, $\kappa = 1.01$ for  $\overline{\Delta {\cal M}(t)}$ of hard-core bosons, and $a=1.11$, $\kappa=0.50$ for $\overline{\Delta {\cal N}(t)}$. (d),(e) The trace distance $d[{\cal G}(t), {\cal G}_{\rm GGE}]$, Eq.~(\ref{trdistance_t_def}), is shown for the same systems as in (a),(b), respectively. (f) The average value of $d[{\cal G}(t), {\cal G}_{\rm GGE}]$, in the time interval $t \in [10^5,10^6]$, is plotted as a function of the system size $L$. The black solid line is a power-law fit for hard-core bosons for the same lattice sizes as in (c), yielding $a=1.22$ and $\kappa=0.52$. Data taken from Ref.~\cite{wright14}.
}
\label{fig_Wright}
\end{figure}

To exemplify the starkly different behavior of some one-body observables for hard-core bosons and noninteracting spinless fermions, we review results from Ref.~\cite{wright14}. There, Wright {\it et al}~studied the dynamics of systems initially prepared in the ground state of a box with twice as many sites as particles, which was then opened in a larger box with four times as many sites as particles (this kind of protocol is sometimes referred to as a geometric quench \cite{caux10, alba14}). A central quantity in that study was the distance between the instantaneous and the GGE momentum distribution, defined as
\begin{equation} \label{deltaM_def}
\Delta {\cal M}(t) = \frac{\sum_k \left\vert m_k(t) - \langle \hat m_k \rangle_{\rm GGE} \right\vert }{\sum_k \langle \hat m_k \rangle_{\rm GGE}}.
\end{equation}
Figure~\ref{fig_Wright}(b) shows that, at long times, $\Delta{\cal M}(t)$ for hard-core bosons decreases with increasing system size. A finite-size scaling analysis of its average $\overline{\Delta{\cal M}(t)}$ reveals a power-law decrease $\propto1/L$ [see Fig.~\ref{fig_Wright}(c)]. In contrast, one can clearly see in Figs.~\ref{fig_Wright}(a) and~\ref{fig_Wright}(c) that for noninteracting spinless fermions the average at long times does not decrease with increasing system size. One might think that this is the result of the fermionic momentum distribution function relaxing to something different from the GGE prediction. However, this is not the case, one can actually prove that the time average of all one-body observables in a noninteracting fermionic system is exactly equal to the GGE prediction, without any finite-size correction \cite{ziraldo_santoro_13, he13}. 

The proof is straightforward if one uses the fact that the eigenstates of the many-body Hamiltonian are products of single-particle eigenstates \cite{he13}. Projecting $\hat{\rho}(t)=|\psi(t)\rangle\langle \psi(t)|$ onto the one-body sector, the time-evolving one-body density matrix takes the form
\begin{equation}\label{eq:singt}
 \hat{\rho}_{\rm ob}(t)=\sum_{q,q'} c_{qq'} e^{-i(\varepsilon_q-\varepsilon_{q'})t}|q\rangle\langle q'|.
\end{equation}
In the absence of degeneracies in the single-particle spectrum (which is the case for the systems studied in this subsection), the infinite-time average of $\hat{\rho}_{\rm ob}(t)$ can be written as
\begin{equation}
\overline{\hat{\rho}_{\rm ob}(t)}=
\lim_{t'\rightarrow \infty} \frac{1}{t'}
\int^{t'}_0 dt\, \hat{\rho}_{\rm ob}(t)
=\sum_q \langle \hat I_q \rangle_0 |q\rangle\langle q|,
\end{equation}
which is, by construction, the one-body density matrix in the GGE.

In contrast to the momentum distribution function, results for the distance between the instantaneous and the GGE site occupations
\begin{equation} \label{deltaN_def}
\Delta {\cal N}(t) = \frac{\sum_i \left\vert n_i(t) - \langle \hat n_i \rangle_{\rm GGE} \right\vert }{\sum_i \langle \hat n_i \rangle_{\rm GGE}},
\end{equation}
which is the same for hard-core bosons and spinless fermions, show that this quantity does decrease with increasing system size [see the inset of Fig.~\ref{fig_Wright}(a)]. A finite-size scaling in Fig.~\ref{fig_Wright}(c) reveals that, at long times, the average $\overline{\Delta {\cal N}(t)}$ decreases as $1/\sqrt{L}$.

\subsection{Noninteracting vs interacting systems mappable to noninteracting ones}

The results for one-body observables in the previous section, in which two observables exhibited equilibration for hard-core bosons, while only one exhibited it for noninteracting fermions, open the question of whether there will be non-equilibrating one-body observables for hard-core bosons of which we are not aware of. After all, hard-core bosons can be mapped onto noninteracting fermions, for which we already found a non-equilibrating one-body observable. 

In order to answer this question, Wright {\it et al}~studied the dynamics of the trace distance between the instantaneous and the GGE one-body density matrices~\cite{wright14}
\begin{equation} \label{trdistance_t_def}
d[{\cal G}(t), {\cal G}_{\rm GGE}] = \frac{1}{2 N} {\rm Tr} \left[ \sqrt{[{\cal G}(t) - {\cal G}_{\rm GGE}]^2} \right],
\end{equation}
where ${\cal G}(t)$ is the instantaneous one-body density matrix, with elements ${\cal G}_{j,l}(t) = \langle \psi(t) | \hat b_j^\dagger \hat b_l |\psi(t) \rangle$ and $\langle \psi(t) | \hat f_j^\dagger \hat f_l |\psi(t) \rangle$ for hard-core bosons and noninteracting fermions, respectively, and ${\cal G}_{\rm GGE}$ is the one-body density matrix in the GGE, with matrix elements $\langle \hat {\cal G}_{j,l} \rangle_{\rm GGE} = \langle \hat b_j^\dagger \hat b_l \rangle_{\rm GGE}$ and $\langle \hat f_j^\dagger \hat f_l \rangle_{\rm GGE}$.

In Figs.~\ref{fig_Wright}(d) and~\ref{fig_Wright}(e), we plot the time evolution of the trace distance for the same systems as in Figs.~\ref{fig_Wright}(a) and~\ref{fig_Wright}(b). The plots show that while the trace distance for hard-core bosons decreases with increasing system size, the trace distance for fermions is constant. The latter is the result of the unitary dynamics of the fermionic one-body sector (the fermions are noninteracting) \cite{wright14}. The finite-size scaling analysis in Fig.~\ref{fig_Wright}(f) reveals that the time average of $d[{\cal G}(t), {\cal G}_{\rm GGE}]$ for hard-core bosons at long times decreases as $1/\sqrt{L}$. Such a power-law decay provides an upper bound for the decay of the time fluctuations in all one-body bases (not only the site and momentum occupations). It shows that no extensive set of one-body observables exists in the hard-core boson system that will fail to exhibit generalized thermalization. This provides an alternative view to generalized thermalization in a closed system where the bath is usually thought of as the physical region of the system that is traced out. Here, the bath is provided by the $N-1$ particles that have been traced out.

Wright~{\it et al}~also studied hard-core anyons, which interpolate smoothly between hard-core bosons and noninteracting spinless fermions, while still being mappable onto the latter. Remarkably, they showed that hard-core anyons exhibit generalized thermalization in the same manner as hard-core bosons do, namely, the average of $d[{\cal G}(t), {\cal G}_{\rm GGE}]$ at long times vanishes with increasing system size as $1/\sqrt{L}$. This makes apparent that noninteracting fermions are a singular limit in this family of models. They are fundamentally different from the interacting models (hard-core bosons and anyons) that were mapped onto them. This is because, in noninteracting fermionic systems, extensive sets of one-body observables can fail to equilibrate, and, hence, can fail to exhibit generalize thermalization, while no such failure can occur for hard-core bosons and anyons. This is made explicit by the fact that the trace distance $d[{\cal G}(t), {\cal G}_{\rm GGE}]$ for spinless fermions is nonzero and constant in time, while the average of $d[{\cal G}(t), {\cal G}_{\rm GGE}]$ for hard-core bosons (anyons) at long times vanishes with increasing system size.

We should stress that in the quenches discussed in this section neither the initial state nor the final Hamiltonian exhibit translational invariance. Lack of translational invariance is a feature common to the quenches studied in Refs.~\cite{rigol07, rigol06, gramsch12, wright14, cassidy11}, in which the GGE was also shown to predict the equilibrated values of observables in isolated integrable systems. Hence, the success of the GGE is in no way tied to translational invariance. We note that, because of open boundary conditions, the single-particle spectrum of the final Hamiltonian in the quenches discussed here (and in Refs.~\cite{rigol07, rigol06, gramsch12, wright14, cassidy11}) is nondegenerate. Extensive degeneracies in the single-particle spectrum after a quench, which occur when a system is translationally invariant, can be the source of subtleties in the GGE that is needed to describe observables after relaxation following a quench \cite{kollar08}, specially from initial states that are not translationally invariant \cite{Fagotti14}.

\section{Ensembles in the transverse field Ising model} \label{diagonal_gge}

This section is devoted to the study of the paradigmatic transverse field Ising model. We consider different statistical ensembles in finite systems and study generalized thermalization of few-body observables. 

Many works in the last twenty years have studied the dynamics of the transverse field Ising model in finite chains \cite{sachdev97, igloi00, Sengupta04, rossini_silva_09, rossini_susuki_10, igloi11, rieger11}. However, we are not aware of studies that compare results obtained within the diagonal ensemble and the GGE.
Here, we are particularly interested on how the differences between the predictions of those ensembles scale with the system size. We also report results for the predictions of the grand canonical ensemble~(\ref{ge_xy}).

We study quenches from the ground state at the transverse field $h_0$, to the final field $h$.
Most of the ingredients needed to calculate the quantities discussed in this section were presented in Sec.~\ref{models}. In some cases, the comparison between the diagonal ensemble and the GGE can be done at the level of analytic expressions. These are possible due to the noninteracting nature of the Hamiltonian~(\ref{Hxy_diag}).

\subsection{Energy distributions} \label{properties_stat}

\begin{figure}[!t]
\includegraphics[width=0.99\textwidth]{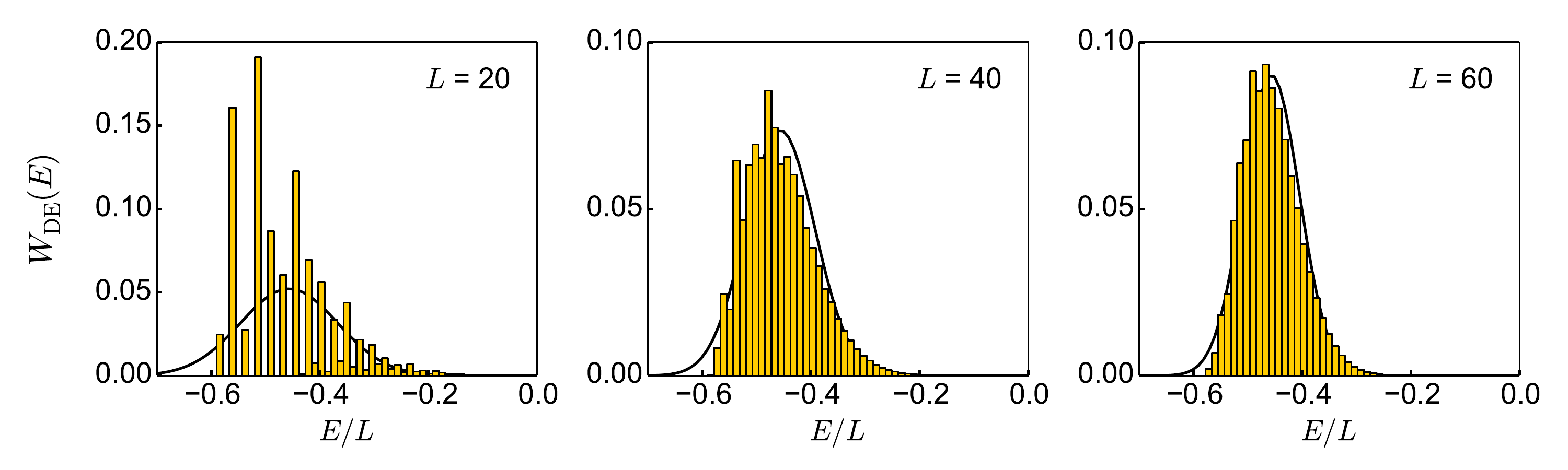}
\caption{
{\it Histogram of the energy distribution in the diagonal ensemble $W_{\rm DE}(E)$ for the transverse field Ising model.}
We quench the transverse field from $h_0=4.0$ to $h=0.8$. Solid lines are Gaussian functions $g(E) = \varepsilon_E /(\sqrt{2 \pi} \sigma) e^{-(E-\bar E)^2/(2\sigma^2)}$ with the mean energy $\bar E = \langle \hat H \rangle_{\rm DE}/L = -0.457$ (the ground-state energy for $h=0.8$ is $E_{\rm gs}/L=-0.584$). We set the bin width $\varepsilon_E = 2|E_{\rm gs}|/100$. The width of $g(E)$ is $\sigma =\sigma_{\hat H,{\rm DE}}/L$, where $\sigma_{\hat H,{\rm DE}} = 0.4\sqrt{L}$ according to Eq.~(\ref{sigma_de}).
}
\label{fig_Edist_DE_080}
\end{figure}

\begin{figure}[!t]
\includegraphics[width=0.99\textwidth]{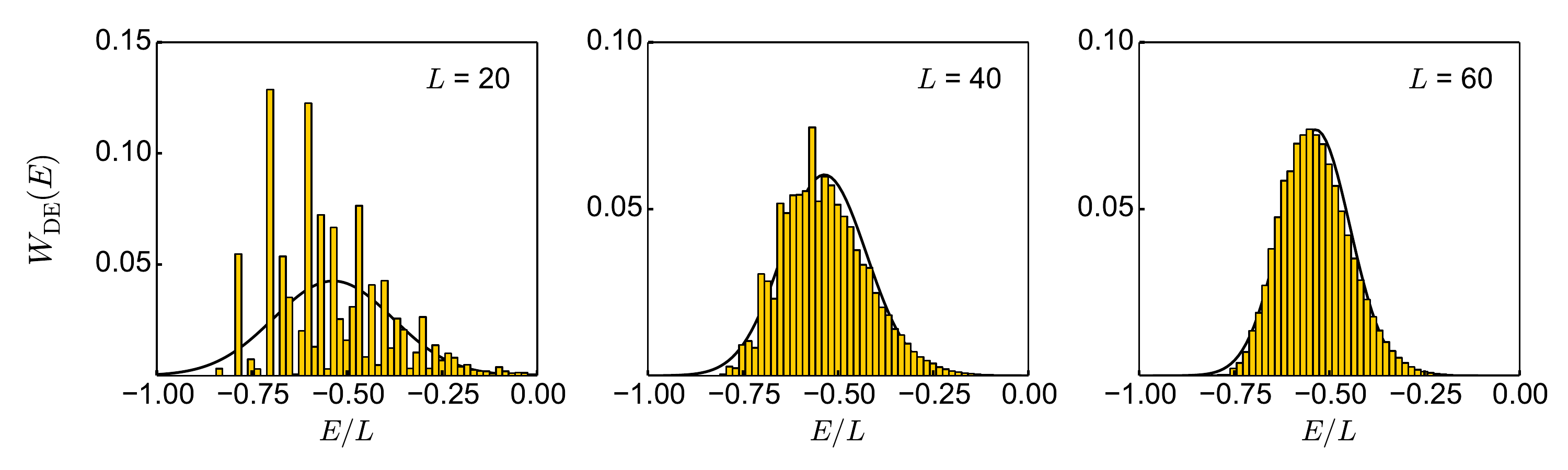}
\caption{
{\it Histogram of the energy distribution in the diagonal ensemble $W_{\rm DE}(E)$ for the transverse field Ising model.}
We quench the transverse field from $h_0=0.1$ to $h=1.5$. Solid lines are Gaussian functions $g(E) = \varepsilon_E /(\sqrt{2 \pi} \sigma) e^{-(E-\bar E)^2/(2\sigma^2)}$ with the mean energy $\bar E = \langle \hat H \rangle_{\rm DE}/L = -0.536$ (the ground-state energy for $h=1.5$ is $E_{\rm gs}/L=-0.836$). We set the bin width $\varepsilon_E = 2|E_{\rm gs}|/100$. The width of $g(E)$ is $\sigma =\sigma_{\hat H,{\rm DE}}/L$, where $\sigma_{\hat H,{\rm DE}} = 0.7\sqrt{L}$ according to Eq.~(\ref{sigma_de}).
}
\label{fig_Edist_DE_150}
\end{figure}

We start by analyzing the decomposition of statistical weights in the energy eigenbasis, i.e., the so-called energy distribution.
In the diagonal ensemble, the energy distribution is defined as
\begin{equation}
W_{\rm DE}(E) = \sum_n |c_n|^2 \delta(E-E_n),
\end{equation}
where $E_n$ are the eigenenergies of the final Hamiltonian and $|c_n|^2$ are the weights of the initial state in the eigenstates of the final Hamiltonian [see Eq.~(\ref{eq:dynam1})]. Coarse grained results for $W_{\rm DE}(E)$ are shown in Fig.~\ref{fig_Edist_DE_080}, for  quenches from the paramagnetic ground state to the ferromagnetic regime, and in Fig.~\ref{fig_Edist_DE_150}, for quenches from the ferromagnetic ground state to the paramagnetic regime. In both cases, we consider chains with three different sizes. Our results show that, with increasing system size, $W_{\rm DE}(E)$ can be well approximated by a Gaussian function with the mean energy and the energy width of the diagonal ensemble.

The mean energy is calculated using Eq.~(\ref{Osingle}) and the overlaps from Eq.~(\ref{eq:overde})
\begin{equation}
\langle \hat H \rangle_{\rm DE} = \sum_{k \in {\cal K}^{(+)}} [\alpha_k (-\varepsilon_k) + (1-\alpha_k) \varepsilon_k] = - \sum_{k \in {\cal K}^{(+)}} (2\alpha_k - 1) \varepsilon_k,
\end{equation}
where $\varepsilon_k$ is the single-particle energy~(\ref{esingle}) and $\alpha_k$ has been defined in Eq.~(\ref{def_alphak}).
The width is computed from
\begin{equation}
\sigma_{\hat H,{\rm DE}}^2 = \langle \hat H^2 \rangle_{\rm DE}  - \langle \hat H \rangle_{\rm DE}^2.
\end{equation}
Since $\langle n | \hat H^2 | n \rangle = \langle n | \hat H | n \rangle \langle n | \hat H | n \rangle$ in the energy eigenbasis, using Eq.~(\ref{O2single}), it follows that
\begin{equation}
\hspace*{-2.5cm}
\langle \hat H^2 \rangle_{\rm DE} = \langle \hat H \rangle_{\rm DE}^2
- \sum_{k \in {\cal K}^{(+)}} \left[ \alpha_k (-\varepsilon_k) + (1-\alpha_k) \varepsilon_k \right]^2
+ \sum_{k \in {\cal K}^{(+)}}  \left[ \alpha_k (-\varepsilon_k)^2 + (1-\alpha_k) \varepsilon_k^2 \right],
\end{equation}
which results in
\begin{equation} \label{H2de}
\sigma_{\hat H,\rm DE}^2 = \sum_{k\in {\cal K}^{(+)}} 4 \alpha_k (1-\alpha_k) \varepsilon_k^2.
\end{equation}

Furthermore, by inserting the expressions for $\varepsilon_k$ and $\alpha_k$ in Eq.~(\ref{H2de}) and taking the continuum limit, one obtains that the width of the energy distribution after the quench is
\begin{equation} \label{sigma_de}
\frac{\sigma_{\hat H,\rm DE}}{\sqrt{L}} =  \left\{ 
\begin{array}{ll}
\frac{1}{2} \left\vert 1- \frac{h}{h_0} \right\vert & \mbox{if } h_0 > 1 \\ \frac{1}{2} h_0 \left\vert 1- \frac{h}{h_0} \right\vert   & \mbox{if } h_0 < 1  
\end{array} \right. .
\end{equation}
We then see that the width is maximal for quenches $(h_0 = 0) \to (h \rightarrow \infty)$, and vice versa. Such quenches result in $\langle \hat H \rangle_{\rm DE} = 0$, which, in turn, corresponds to the energy in the grand canonical ensemble at infinite temperature. In the latter ensemble, one can calculate the width $\sigma_{\hat H, \infty}\equiv\sigma_{\hat H,{\rm GE}}(T\to \infty)$ analytically to get
\begin{equation} \label{sigmaTinf}
\frac{\sigma_{\hat H,\infty}}{\sqrt{L}} =  \frac{\sqrt{1+h^2}}{2}.
\end{equation}
This width agrees with the one in the limiting cases in the diagonal ensemble in which the field is quenched from zero to infinity or vice versa. A similar agreement between the widths in the diagonal ensemble and in the canonical ensemble was discussed in the Bose-Hubbard model for quenches of the onsite repulsion from infinite to zero~\cite{sorg14}.

By performing an analysis for the GGE similar to the one carried out for the diagonal ensemble, one finds that 
\begin{equation}
\hspace*{-2.5cm}
\langle \hat H^2 \rangle_{\rm GGE} = \langle \hat H \rangle_{\rm GGE}^2
- \sum_{k \in {\cal K}^{(+)}} \left[ \alpha_k^2 (-\varepsilon_k) + (1-\alpha_k)^2 \varepsilon_k \right]^2
+ \sum_{k \in {\cal K}^{(+)}}  \left[ \alpha_k^2 (-\varepsilon_k)^2 + (1-\alpha_k)^2 \varepsilon_k^2 \right],
\end{equation}
so that $\sigma_{ \hat H, \rm GGE}^2 = \sum_{k\in {\cal K}^{(+)}} 2 \alpha_k (1-\alpha_k) \varepsilon_k^2$. Therefore, the ratio of the widths of the energy distribution in the diagonal ensemble and the GGE is 
\begin{equation} \label{sigma_de_gge}
\frac{\sigma_{\hat H,\rm DE}}{\sigma_{\hat H,\rm GGE}} = \sqrt{2},
\end{equation}
irrespective of the system size. This result can be intuitively understood already at the level of occupancies within a single $\{k,-k\}$ subspace. The diagonal ensemble only contains states with either both $k$-states empty or occupied (these states have energies $-\varepsilon_k$ and $\varepsilon_k$), while in the GGE all four states, including the ones with zero energy, have nonzero weight. By taking into account the appropriate weights [see Eqs.~(\ref{eq:overde}) and (\ref{eq:overGGE})] in both ensembles, one already recovers the factor $\sqrt{2}$ in Eq.~(\ref{sigma_de_gge}).

\begin{figure}[!t]
\includegraphics[width=0.99\textwidth]{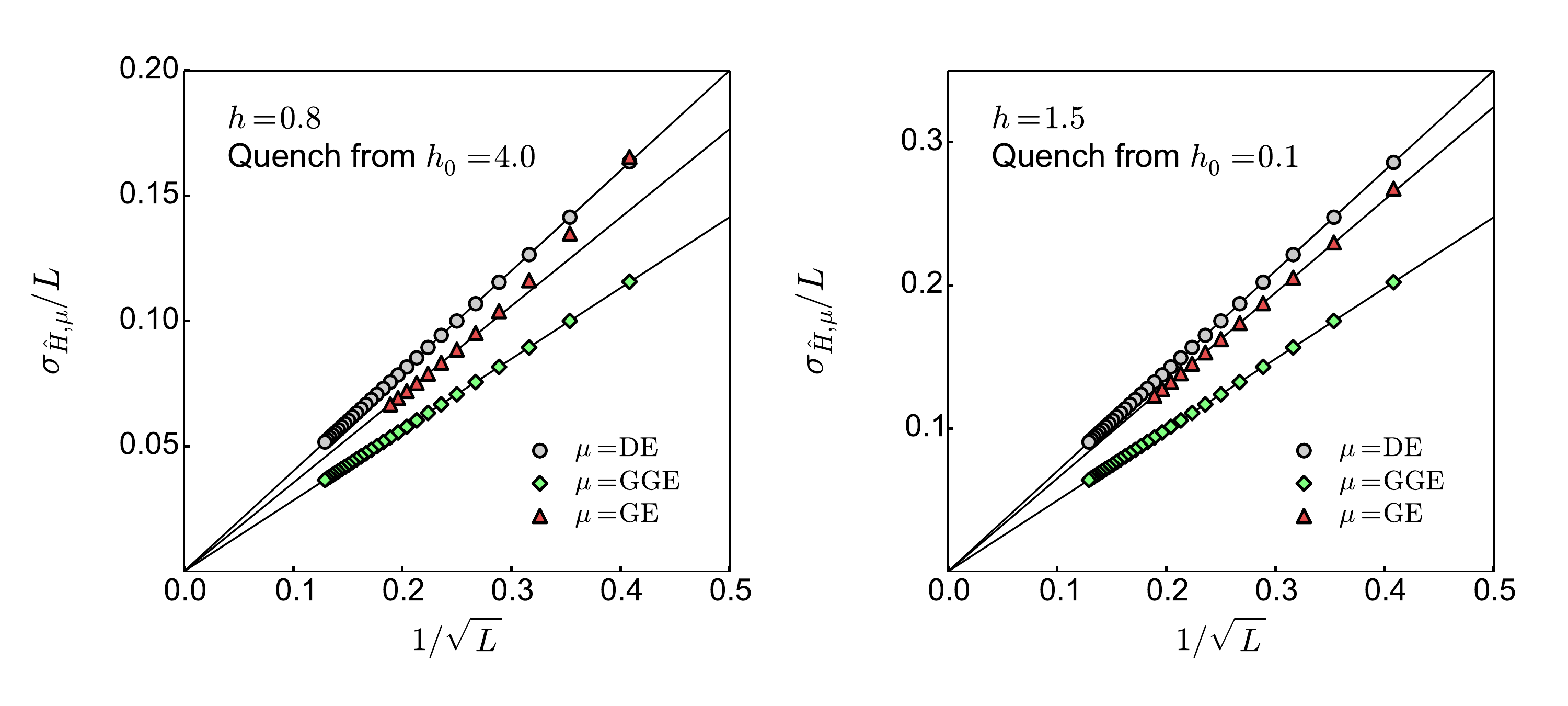}
\caption{
{\it Width of the energy distribution in the GGE, the grand canonical ensemble, and the diagonal ensemble.}
Numerical results (symbols) are presented for quenches from $h_0 = 4.0$ to $h=0.8$ (left panel) and $h_0 = 0.1$ to $h=1.5$ (right panel), and are plotted as a function of $1/\sqrt{L}$. The lines following the results for the diagonal ensemble are $\sigma_{\hat H,{\rm DE}} = 0.4\sqrt{L}$ (left panel) and $\sigma_{\hat H,{\rm DE}} = 0.7\sqrt{L}$ (right panel), see Eq.~(\ref{sigma_de}). The lines following the results for the grand canonical ensemble are linear fits for $L\geq 16$, $\sigma_{\hat H,{\rm GE}} = c_{\rm GE} \sqrt{L}$ with $c_{\rm GE} = 0.353$ (left panel) and $c_{\rm GE} = 0.649$ (right panel). For the GGE, $\sigma_{\hat H,\rm DE}/\sigma_{\hat H,\rm GGE} = \sqrt{2}$ is an exact result (see the derivation of Eq.~(\ref{sigma_de_gge})).
}
\label{fig_Esigma_080_150}
\end{figure}

In Fig.~\ref{fig_Esigma_080_150}, we compare the scaling of the width of the energy distribution in the diagonal ensemble, the GGE, and in the corresponding grand canonical ensemble. In the latter, we calculated the temperature by computing the trace in Eq.~(\ref{ge_xy}) exactly, i.e., by taking the appropriate the sets of wave vectors ${\cal K}^{(+)}$ and ${\cal K}^{(-)}$ in the even and odd sectors, respectively. We report results for two quenches, one from the paramagnetic ground state to the ferromagnetic regime (left panel) and one from the ferromagnetic ground state to the paramagnetic regime (right panel). In all cases under consideration, $\sigma_{\hat H,\rm GE}$ is smaller than $\sigma_{\hat H,\rm DE}$, but larger than $\sigma_{\hat H,\rm GGE}$. Figure~\ref{fig_Esigma_080_150} also shows that, as predicted for the diagonal ensemble and as expected for the grand canonical one, the energy widths vanish as $1/\sqrt{L}$ with increasing system size. That scaling of the widths is generic in quantum quenches within local Hamiltonians~\cite{rigol08}, no matter whether they are integrable or not. 

\begin{figure}[!t]
\includegraphics[width=0.99\textwidth]{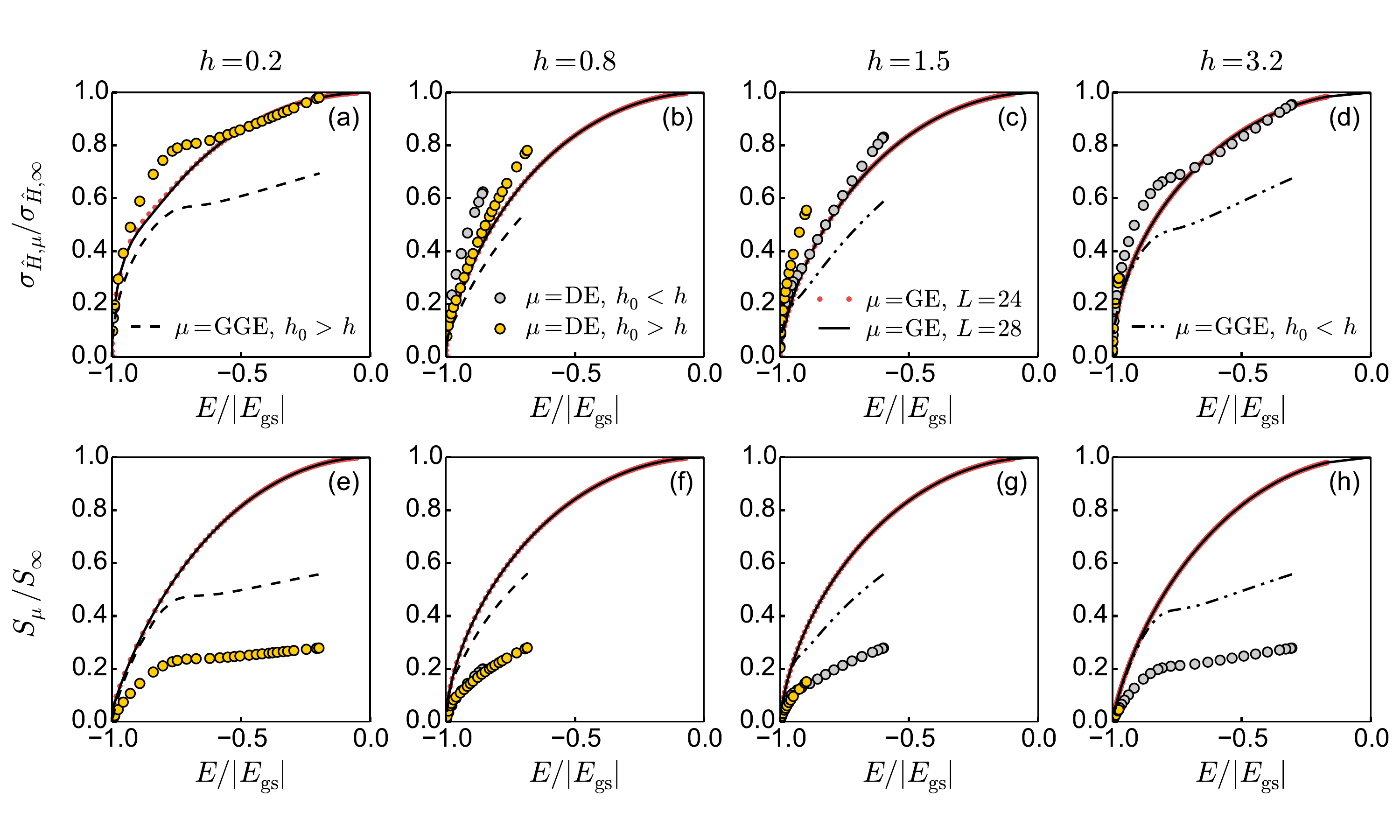}
\caption{
{\it Width of the energy distribution and entropy in the GGE, the grand canonical ensemble, and the diagonal ensemble.} (a)-(d)  $\sigma_{\hat H,{\rm DE}}$, $\sigma_{\hat H,{\rm GGE}}$ and $\sigma_{\hat H,{\rm GE}}$, (e)-(h) $S_{\rm DE}$, $S_{\rm GGE}$ and $S_{\rm GE}$, vs the mean energy (relative to the ground state) after quenches for many different values of $h_0$. Results are reported for four values of the field $h$ after the quench (each column). For the diagonal ensemble, the values of the field $h_0$ ranged from $0.001$ to $5000$.
For the GGE, $\sigma_{\hat H,\rm DE}/\sigma_{\hat H,\rm GGE} = \sqrt{2}$ and $S_{\rm GGE}/S_{\rm DE} = 2$. Black dashed lines correspond to quenches from $h_0>h$ [panels (a)-(b) and (e)-(f), $h=0.2$ and $h=0.8$], and black dotted-dashed lines to quenches from $h_0<h$ [panels (c)-(d) and (g)-(h), $h=1.5$ and $h=3.2$]. For the grand canonical ensemble, the solid lines are the exact results for $L=28$. Red dotted lines, which in almost all cases overlap with solid lines, are shown for comparison and depict results for $L=24$. All results are normalized to the $T=\infty$ values $\sigma_{\hat H,\infty} = \sqrt{1+h^2}/2$, see Eq.~(\ref{sigmaTinf}), and $S_{\infty} = k_{\rm B} \log{\cal D}$.
}
\label{fig_width_entropy}
\end{figure}

In Fig.~\ref{fig_width_entropy}(a)--\ref{fig_width_entropy}(d), we plot $\sigma_{\hat H,{\rm DE}}$, $\sigma_{\hat H,{\rm GGE}}$ and $\sigma_{\hat H,{\rm GE}}$ (rescaled by the infinite temperature result) for four different values of the final field $h$, and for an extended range of initial fields $h_0$. These results provide an overview for when the widths of the diagonal and grand canonical ensembles are close to each other and when they are significantly different. In addition, Figs.~\ref{fig_width_entropy}(b) and~\ref{fig_width_entropy}(c) demonstrate how quenches that lead to the same mean energy can result in markedly different widths of the underlying diagonal ensemble (and, hence, of the GGE). The energy distributions $W_{\rm DE}(E)$ in Figs.~\ref{fig_Edist_DE_080} and~\ref{fig_Edist_DE_150}, which are well fitted by a Gaussian function, may lead one to incorrectly conclude that the initial states considered sample ``ergodically'' the eigenstates of the final Hamiltonian and may result in true thermalization. The fact that this does not happen is what we discuss in the next sections.

\subsection{Entropies}

The results in the previous subsection make apparent that, by looking at coarse grained energy distributions, it might not be possible to differentiate an integrable from a nonintegrable system after a quench. This is where entropy calculations, in particular the calculation of the entropy in the diagonal ensemble \cite{polkovnikov_11, santos_polkovnikov_11}, make a difference. The entropies in the GGE, diagonal, and grand canonical ensembles are computed as 
\begin{equation} \label{Sdef}
S_\mu = -k_{\rm B} {\rm Tr} [\hat \rho_\mu \log \hat \rho_\mu ],
\end{equation}
where $\mu=$ GGE, DE, and GE, respectively. 

Since the weights in all the ensembles factorize, Eq.~(\ref{Sdef}) has the same structure as the expectation value of observables $\langle \hat {\cal O}\rangle_\mu$ in Eq.~(\ref{Oformal}). Using Eq.~(\ref{Osingle}) one can therefore express $S_\mu$ as the sum of single-particle contributions 
\begin{equation} \label{Ssingle}
S_\mu = -k_{\rm B} \sum_{k \in {\cal K}^{(+)}} \left( \sum_{\xi=0}^3 \rho_{k,\mu}^{(\xi)} \log \rho_{k,\mu}^{(\xi)} \right).
\end{equation}
The entropies in the diagonal ensemble and the GGE can be evaluated straightforwardly by inserting the weights in Eqs.~(\ref{eq:overde}) and (\ref{eq:overGGE}), respectively, into Eq.~(\ref{Ssingle}). This yields
\begin{equation}
S_{\rm DE} = -k_{\rm B} \sum_{k \in {\cal K}^{(+)}} \left[ \alpha_k \log(\alpha_k) + (1-\alpha_k) \log (1-\alpha_k) \right],
\end{equation}
while in the GGE, one gets
\begin{eqnarray}
\hspace*{-0.5cm} \hspace*{-2.0cm} S_{\rm GGE} & \hspace*{-1.6cm} = &  - k_{\rm B}\sum_{k \in {\cal K}^{(+)}}  \left[ \alpha_k^2 \log(\alpha_k^2) + 2\alpha_k (1-\alpha_k) \log (\alpha_k (1-\alpha_k)) + (1-\alpha_k)^2 \log (1-\alpha_k)^2 \right] \nonumber \\
& \hspace*{-1.6cm}  = & - 2 k_{\rm B} \sum_{k \in {\cal K}^{(+)}} \left[ \alpha_k \log(\alpha_k) + (1-\alpha_k) \log (1-\alpha_k) \right].
\end{eqnarray}
Closely related to what we found for the energy widths, the ratio
\begin{equation} \label{SdeSgge}
\frac{S_{\rm GGE}}{S_{\rm DE}} = 2
\end{equation}
is independent of the system size and also independent of the choice of the initial eigenstate. For an initial eigenstate that is not the ground state, the weights will in general differ from the ones in the expressions above. This occurs when, for a given $\{ k,-k \}$ subspace, the overlap is nonzero only for one state (the case when $r_k \neq r_{-k}$). If that is the case, $\rho_{k,\mu}^{(\xi)}$ can only take values $1$ or $0$ and hence this subspace does not appear neither in $S_{\rm DE}$ nor in $S_{\rm GGE}$, keeping their ratio unchanged. The factor 2 in Eq.~(\ref{SdeSgge}) reflects the fact that the number of states that can have nonzero weight in the GGE is the square of the number of states that can have nonzero weight in the diagonal ensemble. Of course, the latter condition alone does not guarantee the result in Eq.~(\ref{SdeSgge}). One also needs a special structure in the weights in the GGE and the diagonal ensemble. This doubling of the entropy in the GGE when compared to the diagonal ensemble has been discussed before in the context of the transverse field Ising model \cite{gurarie_13, kormos_bucciantini_14} and for impenetrable bosons with contact interactions in one dimension~\cite{collura_kormos_14}.

In Figs.~\ref{fig_width_entropy}(e)-\ref{fig_width_entropy}(h), we show results for the diagonal, grand canonical, and the GGE entropies for the same quenches for which results were shown for the width of the energy distribution in Figs.~\ref{fig_width_entropy}(a)-\ref{fig_width_entropy}(d). The grand canonical entropy was calculated exactly on a finite system taking into account the proper sets of wave-numbers. The results reported were checked to have negligible finite-size effects. Two things to be remarked here is that the entropy in the diagonal ensemble is extensive, and that it exhibits extensive differences with the entropy of the grand canonical ensemble. Hence, the number of states involved in both ensembles increases exponentially with $L$, but their ratio vanishes with increasing $L$, i.e., the diagonal ensemble contains a vanishingly (exponentially) small fraction of the states in the grand canonical one. Recently, evidence has been reported that, for (generic) experimentally relevant initial states, the same happens in the thermodynamic limit in quenches to integrable models that cannot be mapped onto noninteracting ones \cite{rigol_14,rigol_16}.

\subsection{Generalized thermalization of spin correlations} 

\subsubsection{Local observables.}

The simplest local observable is the on-site magnetization $\hat S^z_j$. It was already pointed out in Ref.~\cite{rossini_silva_09} that its expectation value in the GGE equals its long-time expectation value. Here we compare it to the expectation value computed in the diagonal ensemble.

The eigenstate expectation values for this observable equal
\begin{equation} \label{Sznn}
\langle n | \hat S^z_j |n \rangle = \frac{1}{L} \sum_{k} \left[ |v_k|^2 \left(1 - p_k^{[n]} \right) + |u_k|^2 p_k^{[n]} \right] - \frac{1}{2},
\end{equation}
where the sum runs over all wave vectors\footnote{In the odd sector, however, the sum should not include $k=0$ and $k=\pi$, but should be extended to $\langle n | \hat S^z_j |n \rangle  \to\langle n | \hat S^z_j |n \rangle (k\neq0,k\neq\pi) + (2-p_{k=0} -p_{k=\pi})/L$. This is not needed for our calculations.}. The coefficients $v_k$ and $u_k$ have been introduced in the context of Eq.~(\ref{Hxy_diag}) and equal $|v_k|^2 = (1-a_k/\varepsilon_k)/2$ and $|u_k|^2 = (1 + a_k/\varepsilon_k) /2$. Since the system under consideration is translationally invariant, the site index $j$ does not appear on the r.h.s.~of Eq.~(\ref{Sznn}). An important insight we get from Eq.~(\ref{Sznn}) is how to express $\langle n | \hat S^z_j |n \rangle$ as a sum of single-particle contributions. This allows us to calculate the expectation value of $\hat S_j^z$ in both the diagonal ensemble and the GGE using Eq.~(\ref{Osingle}).

In the diagonal ensemble, we get
\begin{equation}
\hspace*{-2.0cm}
\langle \hat S^z_j \rangle_{\rm DE} = \frac{1}{L} \sum_{k\in {\cal K}^{(+)}} 2 \left[ \alpha_k |v_k|^2 + (1-\alpha_k) |u_k|^2 \right] - \frac{1}{2}=- \frac{1}{L} \sum_{k\in {\cal K}^{(+)}} 
\left(2\alpha_k -1 \right) \frac{a_k}{\varepsilon_k},
\end{equation}
and in the GGE, we get
\begin{equation}
\hspace*{-2.0cm}
\langle \hat S^z_j \rangle_{\rm GGE} =   \frac{1}{L} \sum_{k\in {\cal K}^{(+)}}  \left[ 2\alpha_k^2 |v_k|^2 + 2\alpha_k(1-\alpha_k) (|v_k|^2 + |u_k|^2) + 2(1-\alpha_k)^2 |u_k|^2 \right] -  \frac{1}{2}.
\end{equation}
Simplifying the latter expression results in
\begin{equation} \label{Sz_de_gge}
\langle \hat S^z_j \rangle_{\rm GGE} = \langle \hat S^z_j \rangle_{\rm DE},
\end{equation}
independently of the system size. This is an interesting result given that, in finite systems, our results for the GGE are approximate. The grand canonical trace we used did not account for the difference in boundary conditions between the even and odd sectors.

The standard procedure to calculate expectation values of off-diagonal operators $\hat S_j^a \hat S_{j+r}^a$ (with $a=\{x,y,z\}$ and $r \geq 1$) in many-body eigenstates $\{|n\rangle\}$ is to use the fermionic representation and express the Jordan-Wigner transformation~(\ref{jordan_wigner}) in terms of Majorana fermions using $e^{i\pi \hat f^\dagger_l \hat f_l} = \hat F_l^x \hat F_l^y$, where $\hat F_l^x = \hat f_l^\dagger + \hat f_l$ and $\hat F_l^y = \hat f_l^\dagger - \hat f_l$~\cite{lieb61}. This allows one to express the operators $ \hat S_j^a \hat S_{j+r}^a$ in terms of a string of local fermionic operators acting from site $j$ to site $j+r$.
Since the model is quadratic, the application of Wick's theorem further decomposes the expectation values of a string of local operators in products of expectation values of two operators. The central object in the later decomposition is the operator
\begin{equation} \label{Grdef}
\langle n | \hat G(R) |n\rangle \equiv \langle n | \hat F_m^y \hat F_{m+R}^x | n \rangle = \frac{2}{L} \left( \sum_{k \in {\cal K}^{(+)}} {\cal C}_k^{\left(p_k^{[n]},p_{-k}^{[n]} \right)} (R) \right),
\end{equation}
where
\begin{equation} \label{Cknn}
{\cal C}_k^{\left(p_k^{[n]},p_{-k}^{[n]} \right)} (R) = \left\{
\begin{array}{lcll}
{\cal C}_k^{(0)}(R) &=&  {\cal C}_k(R) & {\rm if} \; \; p_k^{[n]} =0,\ p_{-k}^{[n]} = 0 \\
{\cal C}_k^{(1)}(R) &=&  0 & {\rm if} \; \; p_k^{[n]} =1,\ p_{-k}^{[n]} = 0 \\
{\cal C}_k^{(2)}(R) &=&  0 & {\rm if} \; \; p_k^{[n]} =0,\ p_{-k}^{[n]} = 1 \\
{\cal C}_k^{(3)}(R) &=& - {\cal C}_k(R) & {\rm if} \; \; p_k^{[n]} =1,\  p_{-k}^{[n]} = 1 
\end{array}
\right. ,
\end{equation}
and
\begin{equation} \label{Cksingle}
{\cal C}_k(R) = -\frac{a_k}{\varepsilon_k} \cos(k R) + \frac{b_k}{\varepsilon_k} \sin(k R).
\end{equation}
Using Eq.~(\ref{Osingle}), one can then calculate $\hat G(R)$ in the different ensembles to obtain
\begin{equation} \label{GRde}
\langle \hat G(R) \rangle_{\rm DE} = \langle \hat G(R) \rangle_{\rm GGE} = \frac{2}{L} \sum_{k \in {\cal K}^{(+)}} (2\alpha_k - 1) \, {\cal C}_k(R).
\end{equation}
The fact that the expectation value of $\hat G(R)$ is the same in the diagonal ensemble and in the GGE has its root in symmetric structure of this quantity in the $\{k,-k\}$ subspaces [see Eq.~(\ref{Cknn})], which shares similarities with the energy expectation values. It is also interesting to note that
\begin{equation} \label{SzG0}
\langle \hat S_j^z \rangle_{\rm DE/GGE} = \frac{1}{2} \langle \hat G(0) \rangle_{\rm DE/GGE}.
\end{equation}

The structure of the nearest neighbor correlations of the $x$ and $y$ spin components is particularly simple because $\hat S_j^x \hat S_{j+1}^x = \hat G(1)/4$ and $\hat S_j^y \hat S_{j+1}^y = \hat G(-1)/4$. According to Eq.~(\ref{GRde}), this implies that
\begin{eqnarray}
\langle \hat S_j^x \hat S_{j+1}^x \rangle_{\rm DE} & = & \langle \hat S_j^x \hat S_{j+1}^x \rangle_{\rm GGE} \label{SxSx1}, \label{SxSx1def} \\
\langle \hat S_j^y \hat S_{j+1}^y \rangle_{\rm DE} & = & \langle \hat S_j^y \hat S_{j+1}^y \rangle_{\rm GGE}, \label{SySy1def}
\end{eqnarray}
for any system size. As a result of Eqs.~(\ref{Sz_de_gge}) and~(\ref{SxSx1}), not only does the average energy in the diagonal ensemble equal the one in the GGE, but also the expectation values of the individual terms in the transverse field Ising Hamiltonian are the same in both ensembles.

The next-nearest neighbor correlations of the $x$ and $y$ spin components are more involved since the eigenstate expectation values now contain products of two operators, $\langle n | \hat S_j^x \hat S_{j+2}^x | n \rangle = \frac{1}{4} \left( \langle n | \hat G(1) | n \rangle^2 - \langle n | \hat G(0) | n \rangle  \langle n | \hat G(2) | n \rangle  \right)$ and $\langle n | \hat S_j^y \hat S_{j+2}^y | n \rangle = \frac{1}{4} \left( \langle n | \hat G(-1) | n \rangle^2 - \langle n | \hat G(0) | n \rangle  \langle n | \hat G(-2) | n \rangle  \right) $.
Nevertheless, using Eq.~(\ref{O2single}), the expectation values of these operators can still be calculated in polynomial time
\begin{eqnarray} 
\hspace*{-2.0cm}
\langle \hat S_j^x \hat S_{j+2}^x \rangle_\mu & = & \frac{1}{4} \left( \langle \hat G(1)  \rangle_\mu^2 - \langle  \hat G(0)  \rangle_\mu  \langle  \hat G(2)  \rangle_\mu  \right) + \label{SxSx2} \nonumber \\
\label{SxSx2a}
&&+ \left( \frac{1}{L} \right)^2 \sum_{k \in {\cal K}^{(+)}} \left[ {\cal C}_k(1)^2 - {\cal C}_k(0) {\cal C}_k(2) \right]
\left[ \rho_{k,\mu}^{(0)} + \rho_{k,\mu}^{(3)} - \left( \rho_{k,\mu}^{(0)} - \rho_{k,\mu}^{(3)}  \right)^2  \right]  \\
\hspace*{-2.0cm}
\langle \hat S_j^y \hat S_{j+2}^y \rangle_\mu &= &\frac{1}{4} \left( \langle \hat G(-1)  \rangle_\mu^2 - \langle  \hat G(0)  \rangle_\mu  \langle  \hat G(-2)  \rangle_\mu  \right) + \label{SySy2} 
\nonumber \\
\label{SySy2a}
&& + \left( \frac{1}{L} \right)^2 \sum_{k \in {\cal K}^{(+)}} \left[ {\cal C}_k(-1)^2 - {\cal C}_k(0) {\cal C}_k(-2) \right]
\left[ \rho_{k,\mu}^{(0)} + \rho_{k,\mu}^{(3)} - \left( \rho_{k,\mu}^{(0)} - \rho_{k,\mu}^{(3)}  \right)^2  \right]. 
\end{eqnarray}
The second term in both equations is in fact very similar. First, using Eq.~(\ref{Cksingle}), we find
${\cal C}_k(1)^2 - {\cal C}_k(0) {\cal C}_k(2) = {\cal C}_k(-1)^2 - {\cal C}_k(0) {\cal C}_k(-2) = \sin^2k$.
Second, the weights in the diagonal ensemble equal $\rho_{k,{\rm DE}}^{(0)} + \rho_{k,{\rm DE}}^{(3)} - \left( \rho_{k,{\rm DE}}^{(0)} - \rho_{k,{\rm DE}}^{(3)}  \right)^2 = 4\alpha_k (1-\alpha_k)$, which are twice the values in the GGE, where $\rho_{k,{\rm GGE}}^{(0)} + \rho_{k,{\rm GGE}}^{(3)} - \left( \rho_{k,{\rm GGE}}^{(0)} - \rho_{k,{\rm GGE}}^{(3)}  \right)^2 = 2\alpha_k (1-\alpha_k)$. In addition, the expectation values of $\hat G(R)$ are the same in both ensembles.

We can then simplify Eqs.~(\ref{SxSx2}) and~(\ref{SySy2}) to
\begin{eqnarray}
\langle \hat S_j^x \hat S_{j+2}^x \rangle_{\rm DE} & = & \frac{1}{4} \left( \langle \hat G(1)  \rangle_{\rm DE}^2 - \langle  \hat G(0)  \rangle_{\rm DE}  \langle  \hat G(2)  \rangle_{\rm DE}  \right) 
+ \frac{1}{L} {\cal Z}(1) \label{SxSx2de} \\ 
\langle \hat S_j^x \hat S_{j+2}^x \rangle_{\rm GGE} & = & \frac{1}{4} \left( \langle \hat G(1)  \rangle_{\rm DE}^2 - \langle  \hat G(0)  \rangle_{\rm DE}  \langle  \hat G(2)  \rangle_{\rm DE}  \right) 
+ \frac{1}{2} \frac{1}{L} {\cal Z}(1) \label{SxSx2gge}
\end{eqnarray}
for the $x$ spin component, and in the same manner for the $y$ spin component as
\begin{eqnarray}
\langle \hat S_j^y \hat S_{j+2}^y \rangle_{\rm DE} & = & \frac{1}{4} \left( \langle \hat G(-1)  \rangle_{\rm DE}^2 - \langle  \hat G(0)  \rangle_{\rm DE}  \langle  \hat G(-2)  \rangle_{\rm DE}  \right) 
+ \frac{1}{L} {\cal Z}(1) \label{SySy2de} \\ 
\langle \hat S_j^y \hat S_{j+2}^y \rangle_{\rm GGE} & = & \frac{1}{4} \left( \langle \hat G(-1)  \rangle_{\rm DE}^2 - \langle  \hat G(0)  \rangle_{\rm DE}  \langle  \hat G(-2)  \rangle_{\rm DE}  \right) 
+ \frac{1}{2} \frac{1}{L} {\cal Z}(1). \label{SySy2gge}
\end{eqnarray}
The term ${\cal Z}(1)$ denotes a sum that, more generally, we define as
\begin{equation} \label{Zrdef}
{\cal Z}(r) = \frac{1}{L}  \sum_{k \in {\cal K}^{(+)}} 4 \alpha_k (1-\alpha_k) \sin^2(k r).
\end{equation}
Equations~(\ref{SxSx2de})-(\ref{SxSx2gge}) and~(\ref{SySy2de})-(\ref{SySy2gge}) reveal that the expectation values of $\hat S_j^x \hat S_{j+2}^x $ and $\hat S_j^y \hat S_{j+2}^y$ in the diagonal ensemble and in the GGE are identical in the thermodynamic limit. Their difference for finite systems vanishes as:
\begin{equation}
\langle \hat S_j^x \hat S_{j+2}^x \rangle_{\rm DE} - \langle \hat S_j^x \hat S_{j+2}^x \rangle_{\rm GGE} = 
\langle \hat S_j^y \hat S_{j+2}^y \rangle_{\rm DE} - \langle \hat S_j^y \hat S_{j+2}^y \rangle_{\rm GGE} =  \frac{1}{2L} {\cal Z}(1).
\end{equation}
In the continuum limit, ${\cal Z}(1)$ can be expressed in terms of the transverse fields before and after the quench. For the quenches across the quantum critical point that we consider here, $(h-1)(h_0-1)< 0$, it takes the form
\begin{equation}
{\cal Z}(1) = \frac{| h - h_0 |}{16 \, ({\rm Max}[h,h_0])^3} \left( 3 \, ({\rm Max}[h,h_0])^2 - [1 + h_0 h + (h_0 h)^2] \right).
\end{equation}

\begin{figure}[!t]
\includegraphics[width=0.99\textwidth]{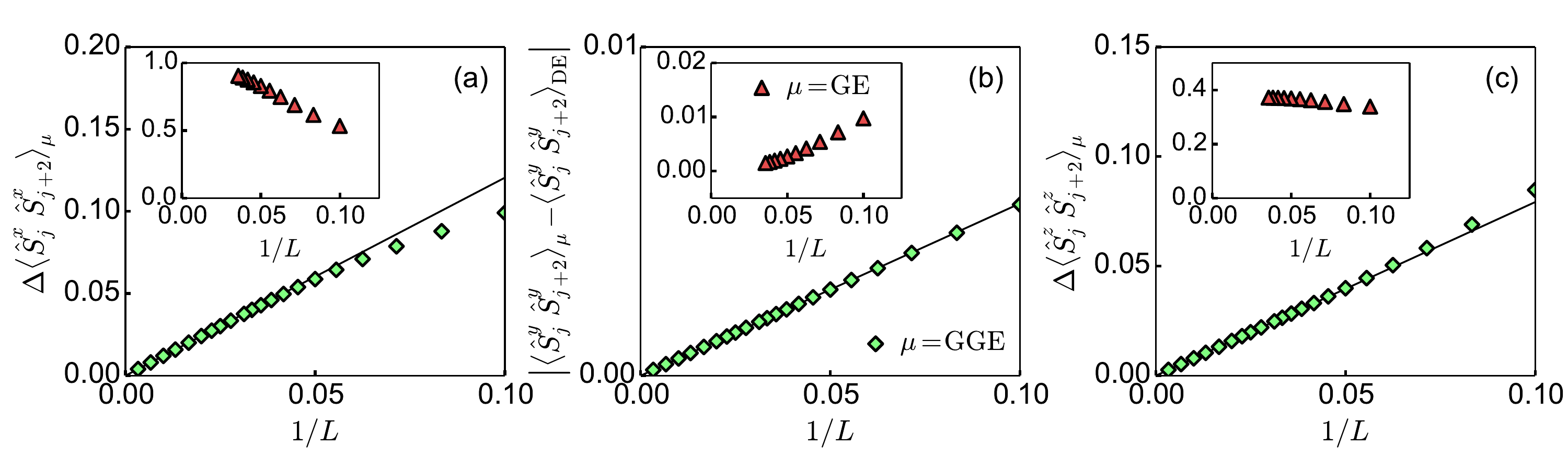}
\caption{
{\it Relative difference between observables in the GGE (main panels), the grand canonical ensemble (insets), and the diagonal ensemble.}
Results are presented for the relative difference $\Delta \langle \hat S_j^a \hat S_{j+2}^a \rangle_\mu$, defined in Eq.~(\ref{DeltaO_def}), for $a=x$ in (a) and $a=z$ in (c). In (b), we plot $| \langle \hat S_j^y \hat S_{j+2}^y \rangle_\mu - \langle \hat S_j^y \hat S_{j+2}^y \rangle_{\rm DE}|$ since the diagonal ensemble result for this observable in the thermodynamic limit is very small, $ \langle \hat S_j^y \hat S_{j+2}^y \rangle_{\rm DE} \approx - 5.7 \times 10^{-4}$. Solid lines are functions $c/L$ for the GGE results, where the coefficient $c$ was obtained from Eqs.~(\ref{SxSx2de}) and~(\ref{SxSx2gge}) for $\hat S_j^x \hat S_{j+2}^x$, Eqs.~(\ref{SySy2de}) and~(\ref{SySy2gge}) for $\hat S_j^y \hat S_{j+2}^y$, and Eqs.~(\ref{SzSzrde}) and~(\ref{SzSzrgge}) for $\hat S_j^z \hat S_{j+2}^z$, replacing sums by integrals. In all quenches the initial state is the ground state in the transverse field $h_0=4.0$ and the final field is $h=0.8$.
}
\label{fig_SaSa2_080}
\end{figure}

\begin{figure}[!t]
\includegraphics[width=0.99\textwidth]{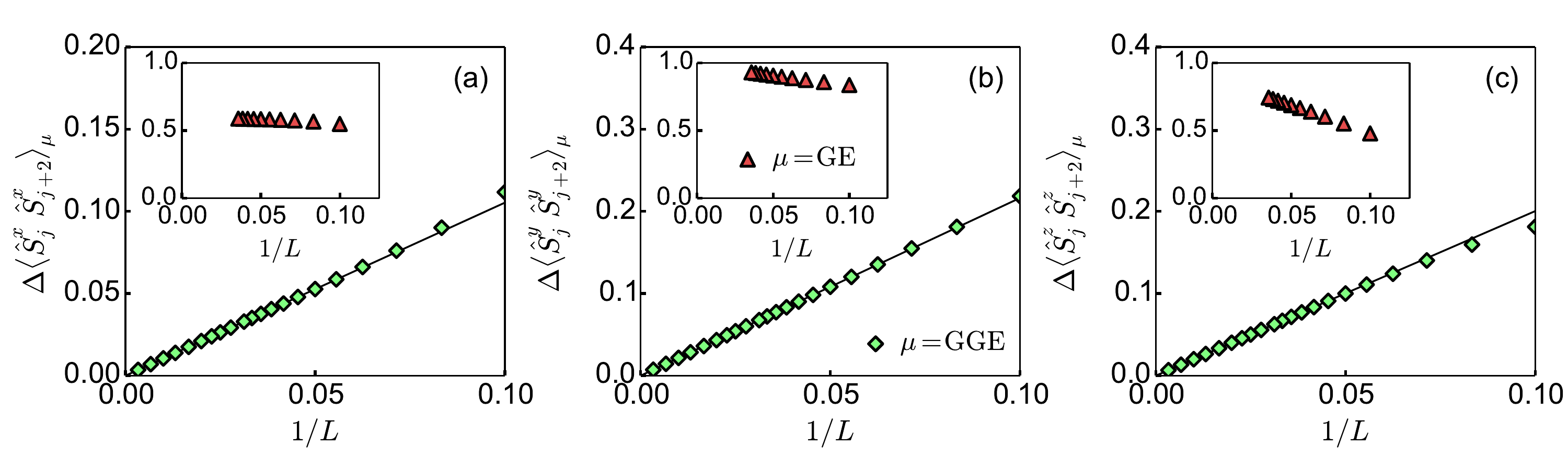}
\caption{
{\it Relative difference between observables in the GGE (main panels), the grand canonical ensemble (insets), and the diagonal ensemble.}
Results are presented for the relative difference $\Delta \langle \hat S_j^a \hat S_{j+2}^a \rangle_\mu$, defined in Eq.~(\ref{DeltaO_def}), for $a=x$ in (a), $a=y$ in (b), and $a=z$ in (c). Solid lines are functions $c/L$ for the GGE results, where the coefficient $c$ was obtained from Eqs.~(\ref{SxSx2de}) and~(\ref{SxSx2gge}) for $\hat S_j^x \hat S_{j+2}^x$, Eqs.~(\ref{SySy2de}) and~(\ref{SySy2gge}) for $\hat S_j^y \hat S_{j+2}^y$ and Eqs.~(\ref{SzSzrde}) and~(\ref{SzSzrgge}) for $\hat S_j^z \hat S_{j+2}^z$, replacing sums by integrals. In all quenches the initial state is the ground state in the transverse field $h_0=0.1$ and the final field is $h=1.5$.
}
\label{fig_SaSa2_150}
\end{figure}

In Figs.~\ref{fig_SaSa2_080}(a)-\ref{fig_SaSa2_080}(b) and Figs.~\ref{fig_SaSa2_150}(a)-\ref{fig_SaSa2_150}(b), we present the numerical results for observables $\hat{\cal O} = \hat S_j^x \hat S_{j+2}^x$ and $\hat{\cal O} = \hat S_j^y \hat S_{j+2}^y$. We plot the relative difference between the predictions of different ensembles as
\begin{equation} \label{DeltaO_def}
\Delta \langle \hat {\cal O} \rangle_\mu = \left\vert \frac{\langle \hat {\cal O} \rangle_\mu - \langle \hat {\cal O} \rangle_{\rm DE}}{ \langle \hat {\cal O} \rangle_{\rm DE}} \right\vert,
\end{equation}
where $\mu=$ GGE and GE. Results are presented for quenches starting in the paramagnetic ground state to the ferromagnetic regime (Fig.~\ref{fig_SaSa2_080}), and for quenches starting in the ferromagnetic ground state to the paramagnetic regime (Fig.~\ref{fig_SaSa2_150}). The agreement between the GGE and the diagonal ensemble, described by Eqs.~(\ref{SxSx2de})-(\ref{SxSx2gge}) and Eqs.~(\ref{SySy2de})-(\ref{SySy2gge}) and depicted in Figs.~\ref{fig_SaSa2_080} and \ref{fig_SaSa2_150}, shows that the observables considered exhibit generalized thermalization in the thermodynamic limit.

In contrast, results for the grand canonical ensemble in the insets in both figures make clear that these systems do not exhibit thermalization, namely, the grand canonical ensemble does not consistently predict expectation values of observables in the diagonal ensemble. We stress that no approximations have been made in the calculation of both the diagonal and the grand canonical ensemble predictions. In the latter case, this means that the eigenstates contain the proper set of wave vectors depending on whether they belong to the even or odd  sector. Even though this limits our numerical calculations of $\langle \hat {\cal O} \rangle_{\rm GE}$ to at most $L\sim 30$ sites, the trend of the data in Figs.~\ref{fig_SaSa2_080} and~\ref{fig_SaSa2_150} is already robust. It makes apparent the failure of traditional statistical mechanics for this system.

Next, we compute the spin correlations in the $z$ direction. An appealing property of these correlations is that their eigenstate expectation values can be expressed as a sum of products of two terms for arbitrary $r$, namely
$\langle n | \hat S_j^z \hat S_{j+r}^z | n \rangle = \frac{1}{4} \left( \langle n | \hat G(0) | n \rangle^2 - \langle n | \hat G(r) | n \rangle  \langle n | \hat G(-r) | n \rangle \right)$. This can be simplified using Eq.~(\ref{O2single}) to obtain
\begin{eqnarray}
\hspace*{-2.0cm}
\langle \hat S_j^z \hat S_{j+r}^z \rangle_\mu & = & \frac{1}{4} \left( \langle \hat G(0)  \rangle_\mu^2 - \langle  \hat G(r)  \rangle_\mu  \langle  \hat G(-r)  \rangle_\mu  \right) + \label{SzSzr} \nonumber \\
&&+ \left( \frac{1}{L} \right)^2 \sum_{k \in {\cal K}^{(+)}} \left[ {\cal C}_k(0)^2 - {\cal C}_k(r) {\cal C}_k(-r) \right]
\left[ \rho_{k,\mu}^{(0)} + \rho_{k,\mu}^{(3)} - \left( \rho_{k,\mu}^{(0)} - \rho_{k,\mu}^{(3)}  \right)^2  \right].  
\end{eqnarray}
In addition, Eq.~(\ref{Cksingle}) implies that ${\cal C}_k(0)^2 - {\cal C}_k(r) {\cal C}_k(-r) = \sin^2(kr)$ and Eq.~(\ref{SzG0}) gives $\langle \hat S_j^z \rangle_{\rm DE} = \frac{1}{2} \langle \hat G(0) \rangle_{\rm DE}$. One can therefore rewrite Eq.~(\ref{SzSzr}) as
\begin{eqnarray}
\langle \hat S_j^z \hat S_{j+r}^z \rangle_{\rm DE} & = & \langle \hat S_j^z \rangle_{\rm DE}^2 - \frac{1}{4}  \langle  \hat G(r)  \rangle_{\rm DE}  \langle  \hat G(-r)  \rangle_{\rm DE} 
+ \frac{1}{L} {\cal Z}(r) \label{SzSzrde} \\ 
\langle \hat S_j^z \hat S_{j+r}^z \rangle_{\rm GGE} & = &  \langle \hat S_j^z \rangle_{\rm DE}^2 - \frac{1}{4}  \langle  \hat G(r)  \rangle_{\rm DE}  \langle  \hat G(-r)  \rangle_{\rm DE}
+ \frac{1}{2} \frac{1}{L} {\cal Z}(r). \label{SzSzrgge}
\end{eqnarray}
This means that the expectation values of spin correlations in the $z$ direction in the GGE and in the diagonal ensemble, for an arbitrary distance $r$, are identical in the thermodynamic limit. Their differences in finite chains vanish as $1/L$, as for $\hat S_j^x \hat S_{j+2}^x$ and $\hat S_j^y \hat S_{j+2}^y$.

In Figs.~\ref{fig_SaSa2_080}(c) and~\ref{fig_SaSa2_150}(c), we plot the relative distance $\Delta \langle \hat S_j^z \hat S_{j+2}^z \rangle_{\mu}$, defined in Eq.~(\ref{DeltaO_def}) (where $\mu = $ GGE, GE), for two sets of quenches. The scaling of $\Delta \langle \hat S_j^z \hat S_{j+2}^z \rangle_{\rm GGE}$ follows the prediction from Eqs.~(\ref{SzSzrde}) and~(\ref{SzSzrgge}). The results for the grand canonical ensemble (inset) show clear differences with respect to the diagonal ensemble and, hence, the inadequacy of the grand canonical ensemble to describe observables after relaxation.

\subsubsection{Trace distances.}

We now turn our focus to spin correlations in the entire system, and ask the question of whether they can be described by the GGE in arbitrary bases (e.g., real space, momentum space, etc). As discussed in Sec.~\ref{relaxation}, the answer to this question has been affirmative for hard-core bosons (and, more generally, for hard-core anyons) in the XX Hamiltonian in the absence of translational invariance~\cite{wright14}. Here we address this question in the context of the translationally invariant transverse field Ising model.

The central object in this calculation is the density matrix ${\cal G}_\mu^a$, with matrix elements $\langle \hat S_j^a \hat S_{l}^a \rangle_\mu $ (for $a=\{ x,y,z \}$). We use Eqs.~(\ref{SzSzrde}) and~(\ref{SzSzrgge}) to calculate the spin correlations in the $z$ direction $\langle \hat S_j^z \hat S_{l}^z \rangle_\mu $. For the spin correlations in the $x$ direction, we calculate the eigenstate expectation values from the Toeplitz matrix~\cite{lieb61}
\begin{equation} \label{SxSxrnn}
\langle n| \hat S_j^x \hat S_{j+r}^x | n \rangle = \frac{1}{4}
\left\vert
\begin{array}{cccc}
\langle n|\hat G(1)|n\rangle & \langle n|\hat G(2)|n\rangle & ... & \langle n|\hat G(r)|n\rangle \\
\vdots &&& \vdots \\
\langle n|\hat G(-r+2)|n\rangle &... & ... & \langle n|\hat G(1)|n\rangle
\end{array}
\right\vert,
\end{equation}
where the matrix elements $\langle n|\hat G(r)|n\rangle$ have been introduced in Eq.~(\ref{Grdef}). Similarly, the eigenstate expectation values of spin correlations in the $y$ direction are obtained as
\begin{equation} \label{SySyrnn}
\langle n| \hat S_j^y \hat S_{j+r}^y | n \rangle = \frac{1}{4}
\left\vert
\begin{array}{cccc}
\langle n|\hat G(-1)|n\rangle & \langle n|\hat G(0)|n\rangle & ... & \langle n|\hat G(r-2)|n\rangle \\
\vdots &&& \vdots \\
\langle n|\hat G(-r)|n\rangle &... & ... & \langle n|\hat G(-1)|n\rangle
\end{array}
\right\vert.
\end{equation}
The eigenstate expectation values from Eqs.~(\ref{SxSxrnn}) and~(\ref{SySyrnn}) are used to calculate expectation values of ${\cal G}_\mu^a$ in statistical ensembles.

\begin{figure}[!t]
\includegraphics[width=0.99\textwidth]{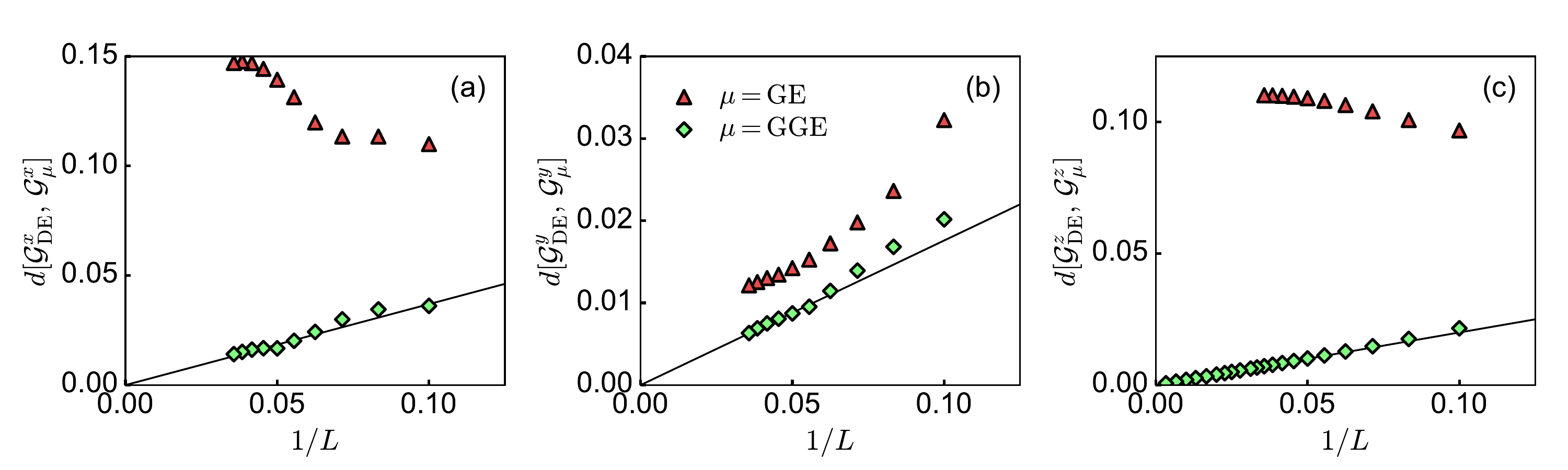}
\caption{
{\it Trace distance between spin correlations in the GGE, or the grand canonical ensemble, and the diagonal ensemble.}
Results are presented for $d[{\cal G}_{\rm DE}^a, {\cal G}_\mu^a]$, defined in Eq.~(\ref{trdistance_def}), for $a=x$ in (a), $a=y$ in (b), and $a=z$ in (c).
Solid lines are fits to power laws $c /L$ for the GGE results.
The corresponding parameters $c = 0.37$ in (a) and $c = 0.18$ in (b) are obtained by fitting the data for $L \geq 18$.
In panel (c), fitting the data for $L \geq 50$ yields $c = 0.20$.
In all quenches the initial state is the ground state for the transverse field $h_0=4.0$ and the final field is $h=0.8$.
}
\label{fig_trDistance_080}
\end{figure}

\begin{figure}[!t]
\includegraphics[width=0.99\textwidth]{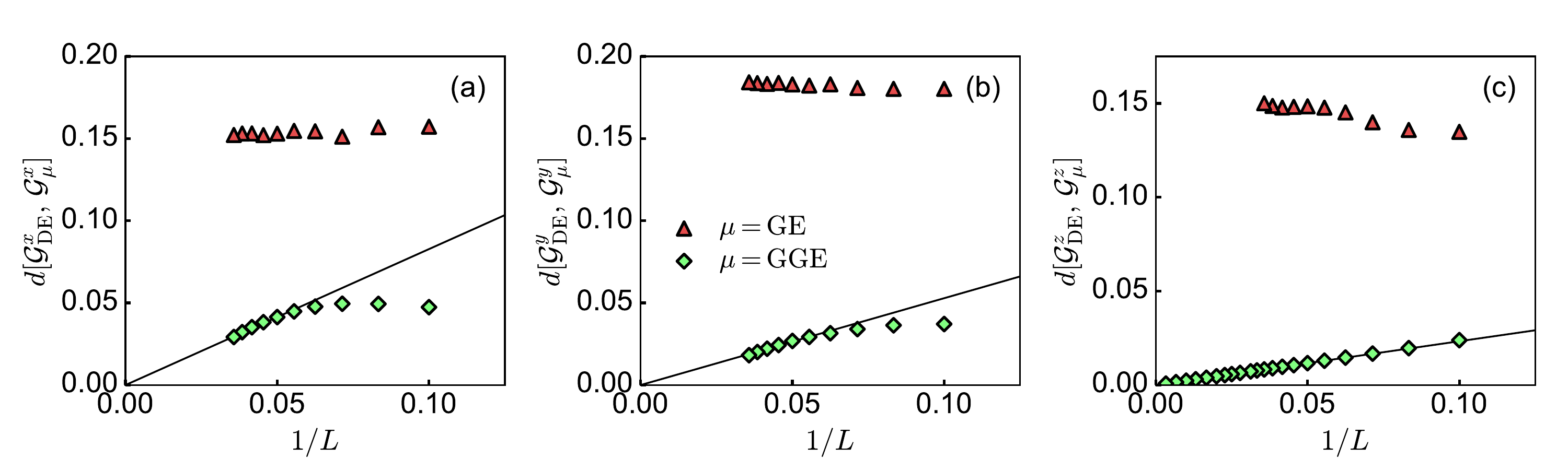}
\caption{
{\it Trace distance between spin correlations in the GGE, or the grand canonical ensemble, and the diagonal ensemble.}
Results are presented for $d[{\cal G}_{\rm DE}^a, {\cal G}_\mu^a]$, defined in Eq.~(\ref{trdistance_def}), for $a=x$ in (a), $a=y$ in (b), and $a=z$ in (c).
Solid lines are fits to power laws $c /L$ for the GGE results.
The corresponding parameters $c = 0.83$ in (a) and $c = 0.53$ in (b) are obtained by fitting the data for $L \geq 18$.
In panel (c), fitting the data for $L \geq 50$ yields $c = 0.23$.
In all quenches the initial state is the ground state for the transverse field $h_0=0.1$ and the final field is $h=1.5$.
}
\label{fig_trDistance_150}
\end{figure}

We define the trace distance between the GGE, or the grand canonical ensemble, and the diagonal ensemble as
\begin{equation} \label{trdistance_def}
d[{\cal G}_{\rm DE}^a, {\cal G}_\mu^a] = \frac{1}{2 {\cal N}} {\rm Tr} \left\{ \sqrt{({\cal G}_{\rm DE}^a - {\cal G}_\mu^a)^2} \right\},
\end{equation}
where $\mu=$ GGE or GE.
In all these cases, the normalization constant (${\cal N}={\rm Tr}\,{\cal G}_\mu^a$) is ${\cal N} = L/4$. In contrast to Eq.~(\ref{trdistance_t_def}), in which we compared instantaneous values to GGE predictions, here we compare predictions of ensembles. Taking into account that our system is translationally invariant, we define $\hat {\cal S}_r^a \equiv \hat  S_j^a \hat  S_{j+r}^a $, whose Fourier transform $\langle \hat  S_k^a\rangle_\mu = (1/L) \sum_{j,l} e^{-i(l-j)k} \langle \hat  S_j^a \hat  S_l^a \rangle_\mu$ simplifies to
\begin{equation}
\langle \hat  S_k^a\rangle_\mu = \frac{1}{4} + 2\sum_{r=1}^{L/2-1} \cos{(kr)} \langle \hat  {\cal S}_r^a\rangle_\mu +  e^{-ikL/2}\langle\hat  {\cal S}_{L/2}^a \rangle_\mu
\end{equation}
and the trace distance can be calculated as
\begin{equation}
d[{\cal G}_{\rm DE}^a, {\cal G}_\mu^a] = \frac{1}{2 {\cal N}} \sum_k \left\vert \langle \hat  S_k^a\rangle_{\rm DE} - \langle \hat  S_k^a\rangle_\mu   \right\vert.
\end{equation}

A particularly simple expression can be obtained for the trace distance of spin correlations in the $z$ direction. In this case, Eqs.~(\ref{SzSzrde}) and~(\ref{SzSzrgge}) imply that 
\begin{equation}
\langle \hat  S_k^z\rangle_{\rm DE} - \langle \hat  S_k^z\rangle_{\rm GGE} = \frac{1}{2L} \left[
 2\sum_{r=1}^{L/2-1} \cos{(kr)} {\cal Z}(r) +  e^{-ikL/2} {\cal Z}(L/2)
 \right].
\end{equation}
Hence, we need to evaluate ${\cal Z}(r)$, introduced in Eq.~(\ref{Zrdef}). In the continuum limit, one can obtain a closed expression for ${\cal Z}(r)$. For quenches across the critical point, $(h-1)(h_0-1) < 0$, it reads
\begin{eqnarray}
\hspace*{-1.0cm}
{\cal Z}(r) &=& \frac{| h - h_0 |}{16 \, (1-h_0h)} \Bigg( ({\rm Max}[h,h_0])^{-(2r+1)} [({\rm Max}[h,h_0])^2-1] \\ && + ({\rm Min}[h,h_0])^{2r-1} [({\rm Min}[h,h_0])^2-1] + 2 \left[({\rm Max}[h,h_0])^{-1} - {\rm Min}[h,h_0] \right]   \Bigg).\nonumber
\end{eqnarray}
For large $r$, ${\cal Z}(r)$ approaches a constant. We plot the trace distance $d[{\cal G}_{\rm DE}^z, {\cal G}_{\mu}^z]$ in Figs.~\ref{fig_trDistance_080}(c) and~\ref{fig_trDistance_150}(c) for two different quenches. The trace distance between the diagonal ensemble and the GGE decays as a power law $\sim 1/L$, in contrast to the results in the grand canonical ensemble where no decay is observed.

In Figs.~\ref{fig_trDistance_080} and~\ref{fig_trDistance_150}, we also show $d[{\cal G}_{\rm DE}^a, {\cal G}_{\rm GGE}^a]$ and $d[{\cal G}_{\rm DE}^a, {\cal G}_{\rm GE}^a]$ for two different quenches and the spin components $a = \{ x,y \}$. As for nearest neighbor correlations, one can see that $d[{\cal G}_{\rm DE}^a, {\cal G}_{\rm GGE}^a]$ decreases with increasing system size,  suggesting generalized thermalization of spin correlations on the entire lattice, while in most cases $d[{\cal G}_{\rm DE}^a, {\cal G}_{\rm GE}^a]$ increases or saturates. Granted, for trace distances finite size effects are stronger than for nearest neighbor correlations and power law behavior ($\sim 1/L$) in the decrease of the trace distance is only seen for the largest system sizes. Similar scaling of the differences between the diagonal ensemble predictions (or the time-average predictions) and the GGE for momentum distribution functions were found for the XX model in Refs.~\cite{gramsch12, cassidy11}. 

It is important to stress that, after equilibration following a quench in the 1D systems discussed here, the spin correlations (and the one-body correlations for hard-core bosons) generally decay exponentially with the distance. Hence, despite the fact that trace distances for spin correlations (and the momentum distribution functions for hard-core bosons) are nonlocal by definition, they effectively behave as local quantities because of the exponential decay of correlations. These results make apparent that the relation between the locality of observables and the applicability of the GGE is a blurry one, because the operators may be by definition nonlocal, but effectively behave as local.

\section{Generalized Eigenstate Thermalization} \label{geth}

Looking back to the definition of the diagonal ensemble [Eq.~(\ref{de_def})], and comparing it to the definition of the GGE [Eq.~(\ref{gge_def})], one cannot help but wonder why the GGE can describe observables after relaxation. The diagonal ensemble is constructed with an exponentially large (in the system size) number of parameters. They are the projection of the initial state onto the eigenstates of the final Hamiltonian. The GGE, on the other hand, is constructed with a polynomially large (in the system size) number of parameters. Those are the Lagrange multipliers, which are determined in terms of the occupations of the single-particle states used to diagonalize the integrable model after the quench. This means that the diagonal ensemble contains exponentially many more parameters than the GGE. The fact that the diagonal ensemble is more constrained than the GGE is apparent in our results for the entropies in both ensembles within the translationally invariant transverse field Ising model. We have shown that the entropy of the former is one half that of the latter. Numerical results in systems with no translational invariance have found that while the entropy of the GGE is always greater than that of the diagonal ensemble, the ratio between the two need not be 2 \cite{rigol_fitzpatrick_11, santos_polkovnikov_11, he_rigol_12}.

A way to understand how it is possible, in general, that two ensembles sampling vastly different number of states (the ratio between the number of states sampled by the diagonal ensemble and by the GGE vanishes exponentially fast with increasing system size) lead to the same results for observables (up to finite size effects) was put forward in Ref.~\cite{cassidy11}. There it was shown that, in a system of hard-core bosons, eigenstates of the Hamiltonian that have similar distributions of conserved quantities also have similar expectation values of observables (such as the occupation of momentum modes). This phenomenon was named {\it generalized eigenstate thermalization} after a related phenomenon in nonintegrable systems, namely, eigenstate thermalization \cite{deutsch91, srednicki94, rigol08}. A system is said to exhibit eigenstate thermalization if eigenstates with close energies have similar expectation values of observables (with deviations that decrease exponentially with increasing system size \cite{dalessio_kafri_15}). How generalized eigenstate thermalization can explain the success of the GGE, as well as the fact that it occurs in the transverse field Ising model, is something that we explain and exemplify in what follows. We note that related ideas have been discussed in the context of the so-called quench-action method \cite{wouters_denardis_14, pozsgay_mestyan14, ilievski15b, caux_essler13, nardis14} (see also the review by Caux in this volume~\cite{caux16}).

\subsection{The XX model. Hard-core bosons}

Before discussing generalized eigenstate thermalization in the transverse field Ising model, we summarize the main points and results reported in Ref.~\cite{cassidy11} and in a later study \cite{he13} of this phenomenon for hard-core bosons (XX model) with the Hamiltonian~(\ref{Hxx_def}). 

The fact that the total number of particles is conserved in such a model motivated the introduction of a microcanonical version of the GGE in Ref.~\cite{cassidy11}, the generalized microcanonical ensemble\footnote{In has been proved in a recent study that the GME is the ensemble that correctly predicts the time average of observables in isolated classical systems \cite{Yuzbashyan16}.} (GME). A key object in the construction of the GME is the target distribution of conserved quantities, which was chosen to be a coarse gained version of the distribution of conserved quantities in the initial state. An eigenstate of the Hamiltonian belongs to the GME if it has a distribution of the conserved quantities whose distance (see Ref.~\cite{cassidy11} for the definition of distance) to the target distribution is below some threshold, similarly to what one does to construct the microcanonical ensemble. The GME constructed this way was shown to accurately reproduce the expectation values of observables in the diagonal ensemble, in contrast to the traditional microcanonical ensemble, which was shown to fail. Having all the states that participated in the diagonal ensemble and in the GME, it was possible to show that the expectation value of observables in those states had a narrow distribution centered about the mean value predicted by the diagonal ensemble and a standard deviation that vanished with increasing system size, i.e., the exact weights and number of states used in each ensemble was not important \cite{cassidy11}. Results obtained in Ref.~\cite{he13} for hard-core bosons in the presence of a quasi-periodic potential were consistent with these findings, but only in the regime in which single-particle eigenstates were extended in real space, i.e., in the absence of localization.

These results are to be contrasted with what happens if one studies the expectation values of observables in all eigenstates of an integrable Hamiltonian that are within a narrow energy window \cite{cassidy11, ikeda_watanabe_13, alba_15}. In this case one finds that they exhibit a narrow distribution centered about the mean value predicted by traditional statistical mechanics, with a standard deviation that vanishes with increasing system size (as $1/\sqrt{L}$ \cite{alba_15}). This means that the only way in which observables in integrable systems can fail to thermalize after a quench is if the initial state samples a vanishingly small fraction of the states in the microcanonical window. This has been recently argued to be generic in quenches that are experimentally relevant \cite{rigol_16}. 

\subsection{The transverse field Ising model}

\begin{figure}[!t]
\includegraphics[width=0.99\textwidth]{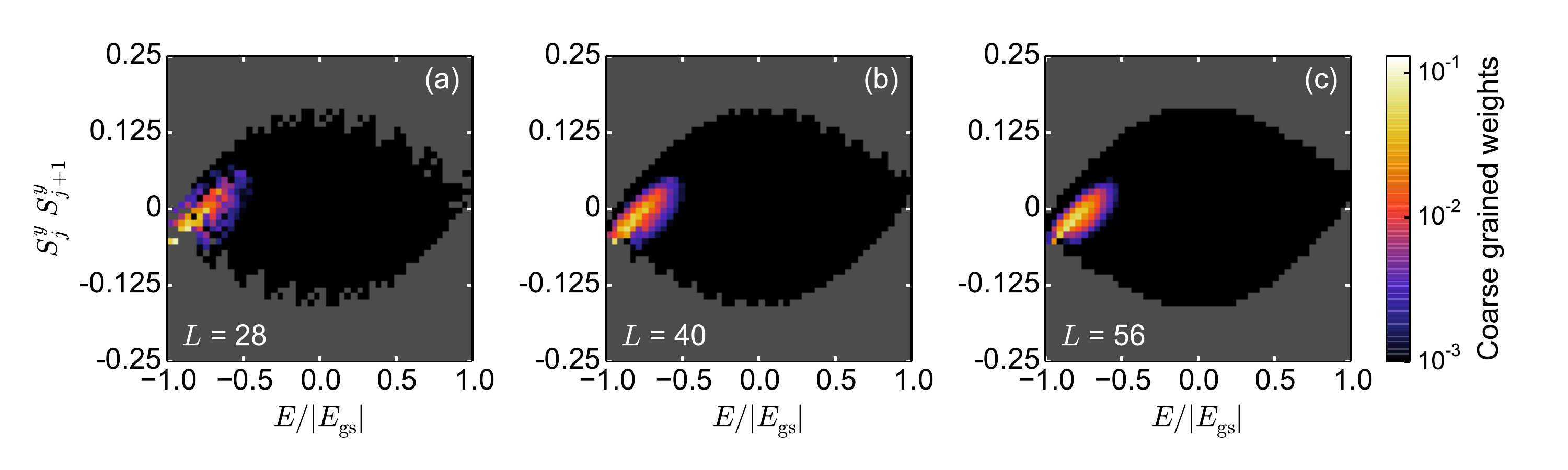} \\
\includegraphics[width=0.99\textwidth]{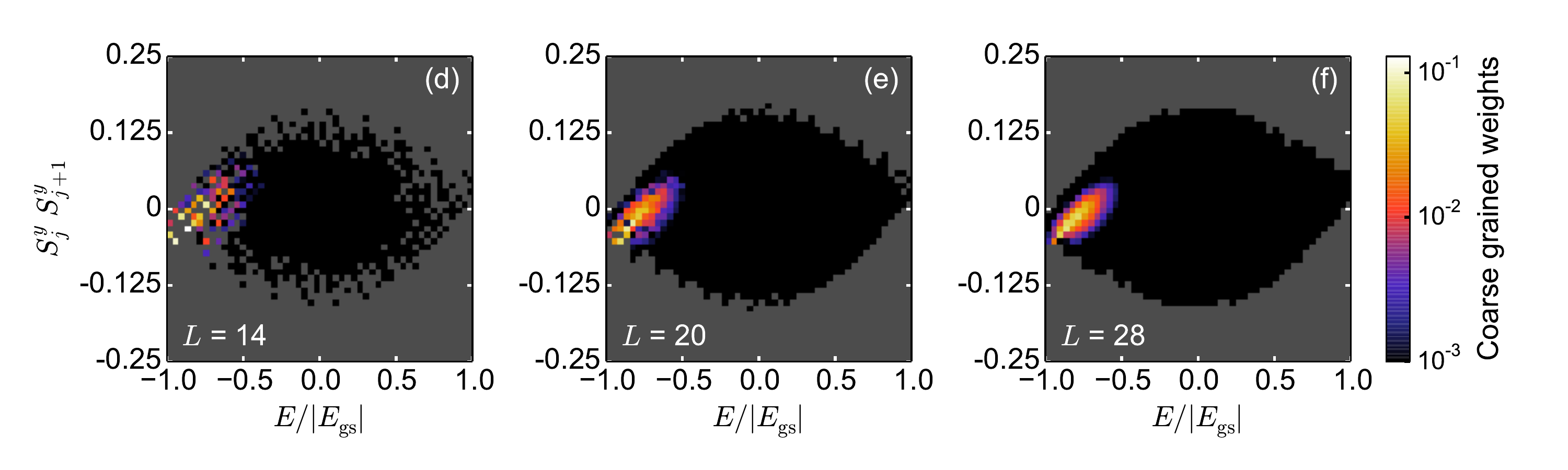}
\caption{
{\it Density plot of the weights of the Hamiltonian eigenstates in terms of their energies and eigenstate expectation values.}
Results are presented for the nearest neighbor $S^y_jS^y_{j+1}$ correlations in the diagonal ensemble [panels (a)-(c)] and in the GGE [panels (d)-(f)]. We quench from the ground state for $h_0=4.0$ to $h=0.8$. Black pixels mark the presence of eigenstates (with vanishing weight), while gray pixels denote their absence. Colored pixels show the nonvanishing weights $\rho^{[n]}_{\rm DE}$ and $\rho^{[n]}_{\rm GGE}$ in the diagonal ensemble and the GGE, respectively. We set the horizontal pixel width to be $\varepsilon_E = 2|E_{\rm gs}|/50$ and the vertical pixel width to be $\varepsilon_{\cal O} = (1/4)\times 2/50$. The ground-state energy for $h=0.8$ is $E_{\rm gs}/L=-0.584$.
}
\label{fig_sysyr1_density_080}
\end{figure}

\begin{figure}[!t]
\includegraphics[width=0.99\textwidth]{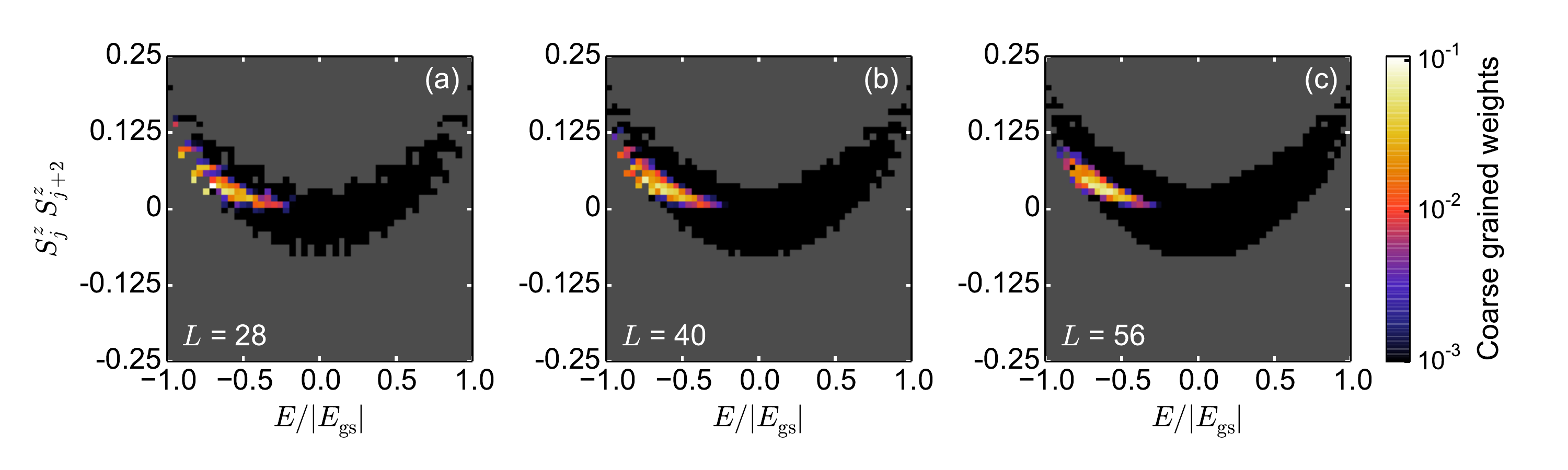} \\
\includegraphics[width=0.99\textwidth]{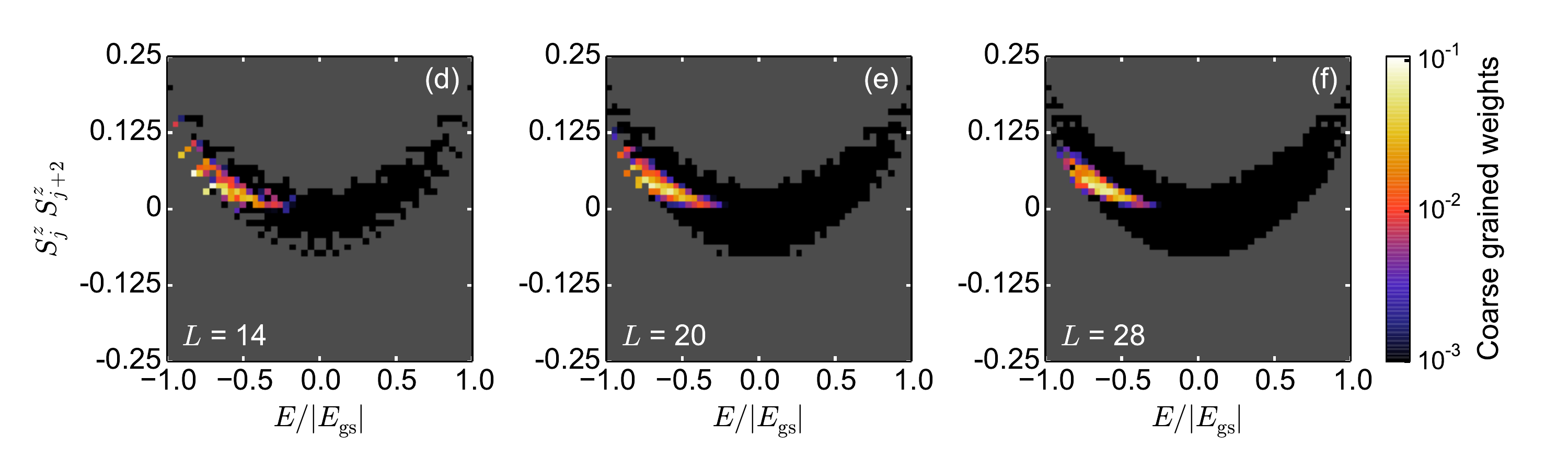}
\caption{
{\it Density plot of the weights of the Hamiltonian eigenstates in terms of their energies and eigenstate expectation values.}
Results are presented for the nearest neighbor $S^z_jS^z_{j+2}$ correlations in the diagonal ensemble [panels (a)-(c)] and in the GGE [panels (d)-(f)]. We quench from the ground state for $h_0=0.1$ to $h=1.5$. Black pixels mark the presence of eigenstates (with vanishing weight), while gray pixels denote their absence. Colored pixels show the nonvanishing weights $\rho^{[n]}_{\rm DE}$ and $\rho^{[n]}_{\rm GGE}$ in the diagonal ensemble and the GGE, respectively. We set the horizontal pixel width to be $\varepsilon_E = 2|E_{\rm gs}|/50$ and the vertical pixel width to be $\varepsilon_{\cal O} = (1/4)\times 2/50$. The ground-state energy for $h=1.5$ is $E_{\rm gs}/L=-0.836$.
}
\label{fig_szszr2_density_150}
\end{figure}

In the hard-core boson language, the transverse field Ising model does not exhibit particle number conservation. This means that only the grand canonical ensemble and the GGE are meaningful for this model. Hence, we carry out a study parallel to the one in Refs.~\cite{cassidy11, he13} but considering grand canonical ensembles instead of microcanonical ones. We compute the eigenstate expectation values of all states in the diagonal ensemble and in the GGE. Since the former involves only the square root of all the eigenstates in the Hilbert space, we can carry out this calculation for twice as many sites as for the latter. We also calculate the weights of all states in both ensembles. 

In Figs.~\ref{fig_sysyr1_density_080} and~\ref{fig_szszr2_density_150}, we show results for two observables and two different quenches. The plots provide a coarse grained view of the exact calculations described above. The top panels in the figures show the results obtained in the diagonal ensemble and the bottom panels show the results obtained in the GGE, for three different system sizes in each case (notice the difference in system sizes between the top and the bottom panels). The black regions mark the existence of eigenstates with those eigenenergies and expectation values of observables, but which have negligible weight in the ensembles. The fact that those black regions are wide and do not narrow with increasing system size make apparent that the transverse field Ising model does not exhibit eigenstate thermalization.

A remarkable feature seen in Figs.~\ref{fig_sysyr1_density_080} and \ref{fig_szszr2_density_150} is that the eigenstates of the Hamiltonian that contribute significantly to the diagonal ensemble and the GGE are narrowly distributed in the same region about a line (notice the log scale used for the coarse grained weights). If one forgets about the other eigenstates (the ones with negligible weights) the behavior seen in Figs.~\ref{fig_sysyr1_density_080} and \ref{fig_szszr2_density_150} is very similar to the one seen for various observables in nonintegrable one-dimensional systems, which exhibit eigenstate thermalization \cite{rigol_09a,rigol_09b}. 

We note that, in Figs.~\ref{fig_sysyr1_density_080} and \ref{fig_szszr2_density_150}, the region with nonvanishing coarse grained weights shrinks in both directions as the system size increases. In the horizontal direction (energy axis), the width decreases as $\sim 1/\sqrt{L}$, as demonstrated in Fig.~\ref{fig_Esigma_080_150}. Hence, here we focus on the variance for the vertical direction
\begin{equation} \label{Sigma_O}
\Sigma_{\hat{\cal O},\mu}^2 = \sum_n  \rho_\mu^{[n]} \langle n | \hat {\cal O} | n \rangle^2  - \left( \sum_n \rho_\mu^{[n]} \langle n | \hat {\cal O} | n \rangle \right)^2,
\end{equation}
where $\mu = {{\rm DE},{\rm GGE}}$. Using Eqs.~(\ref{Osingle}) and (\ref{O2single}), Eq.~(\ref{Sigma_O}) can be  computed straightforwardly for the transverse magnetization $\hat S_j^z$, and the nearest neighbor spin-spin correlations $\hat S_j^x \hat S_{j+1}^x$ and $\hat S_j^y \hat S_{j+1}^y$.

For the transverse magnetization, one can express $\Sigma_{\hat S_j^z,\mu}^2$ as
\begin{eqnarray} \label{SigmaSz1}
\Sigma_{\hat S_j^z,\mu}^2 & = & \frac{1}{4} \left[ \langle \hat G(0) \rangle_\mu \langle \hat G(0) \rangle_\mu
- \left( \frac{2}{L} \right)^2 \sum_{k \in {\cal K}^{(+)}} \left( \sum_{\xi = 0}^3 \rho_{k,\mu}^{(\xi)} {\cal C}_{k}^{(\xi)}(0) \right) \left( \sum_{\xi' = 0}^3 \rho_{k,\mu}^{(\xi')} {\cal C}_{k}^{(\xi')}(0)  \right) \right. \nonumber \\
&& \left. + \left( \frac{2}{L} \right)^2 \sum_{k \in {\cal K}^{(+)}} \left( \sum_{\xi = 0}^3 \rho_{k,\mu}^{(\xi)}[ {\cal C}_{k}^{(\xi)}(0)]^2  \right)
- \langle \hat G(0) \rangle_\mu^2 \right],
\end{eqnarray}
where we have used that $\langle \hat S_j^z \rangle_{\rm DE} = \frac{1}{2} \langle \hat G(0) \rangle_{\rm DE}$, see Eq.~(\ref{SzG0}). A further simplification of Eq.~(\ref{SigmaSz1}) can be achieved by replacing ${\cal C}_{k}^{(\xi)}(0)$ according to Eq.~(\ref{Cknn}), which gives
\begin{equation} \label{SigmaSz2}
\Sigma_{\hat S_j^z,\mu}^2 = \frac{1}{L^2}  \sum_{k \in {\cal K}^{(+)}}  [{\cal C}_{k}(0)]^2  \left[ \rho_{k,\mu}^{(0)} + \rho_{k,\mu}^{(3)} - \left( \rho_{k,\mu}^{(0)} - \rho_{k,\mu}^{(3)}  \right)^2  \right].
\end{equation}
An expression for the sum of the weights above was obtained in the context of Eqs.~(\ref{SxSx2a}) and~(\ref{SySy2a}), and, from Eq.~(\ref{Cksingle}), we get that $[{\cal C}_{k}(0)]^2 = (a_k/\varepsilon_k)^2$. Hence, in the diagonal ensemble, the variance of the distribution of $\hat S_j^z$ equals
\begin{equation} \label{SigmaSzde}
\Sigma_{\hat S_j^z,{\rm DE}}^2 = \frac{1}{L^2}  \sum_{k \in {\cal K}^{(+)}} 4 \alpha_k (1-\alpha_k) \left( \frac{a_k}{\varepsilon_k} \right)^2.
\end{equation}
Equation~(\ref{SigmaSzde}) advances that the width of the distribution vanishes as $\Sigma_{\hat S_j^z,{\rm DE}} \sim 1/\sqrt{L}$ with increasing system size. This scaling is the same as for the width of the energy density distribution derived in Eq.~(\ref{H2de}). In addition, by inserting the weights of the GGE in Eq.~(\ref{SigmaSz2}), one gets
\begin{equation}
\frac{\Sigma_{\hat S_j^z,{\rm DE}}}{\Sigma_{\hat S_j^z,{\rm GGE}}} = \sqrt{2}.
\end{equation}
The ratio above is the same as for the energy distribution~(\ref{sigma_de_gge}). Furthermore, Eq.~(\ref{SigmaSzde}) can be evaluated analytically in the continuum limit. For the quenches across the critical point that we consider in this review, one gets
\begin{equation}
\Sigma_{\hat S_j^z,{\rm DE}}^2 =
\left\{
\begin{array}{lllll}
\frac{1}{4L} \left[ 1+ \frac{h_0 (2+ h_0 h) - h (3 + 4 h_0 h)}{4 h^3} \right] & {\rm if} & h>1 & {\rm and} & h_0 < 1 \\
\frac{1}{16L} \left[ 1+ \frac{1 - 2 h_0 h}{h_0^2} \right] & {\rm if} & h<1 & {\rm and} & h_0 > 1
\end{array} \right. .
\end{equation}
This equation allows one to predict the prefactor $c$ in $\Sigma_{\hat S_j^z,{\rm DE}}  = c/\sqrt{L}$ and its functional dependence on the transverse magnetic fields. It turns out that the largest value of $c$ (hence, the slowest decay of $\Sigma_{\hat S_j^z,{\rm DE}}$) is $c=1/2$, which is realized for quenches in which $h \to \infty$.

It is straightforward now to carry out a similar calculation to determine the widths of $\hat S_j^x \hat S_{j+1}^x$ and $\hat S_j^y \hat S_{j+1}^y$. In analogy to Eq.~(\ref{SigmaSz2}), we get 
\begin{eqnarray}
\Sigma_{\hat S_j^x \hat S_{j+1}^x,\mu}^2 &=& \frac{1}{4L^2}  \sum_{k \in {\cal K}^{(+)}}  [{\cal C}_{k}(1)]^2  \left[ \rho_{k,\mu}^{(0)} + \rho_{k,\mu}^{(3)} - \left( \rho_{k,\mu}^{(0)} - \rho_{k,\mu}^{(3)}  \right)^2  \right] \\
\Sigma_{\hat S_j^y \hat S_{j+1}^y,\mu}^2 &=& \frac{1}{4L^2}  \sum_{k \in {\cal K}^{(+)}}  [{\cal C}_{k}(-1)]^2  \left[ \rho_{k,\mu}^{(0)} + \rho_{k,\mu}^{(3)} - \left( \rho_{k,\mu}^{(0)} - \rho_{k,\mu}^{(3)}  \right)^2  \right],
\end{eqnarray}
where the weights are identical as in Eq.~(\ref{SigmaSz2}). The only difference between the expressions for the different observables is the kernel function ${\cal C}_{k}(r)$ involved. The latter is given by Eq.~(\ref{Cksingle}), and can be simplified for $r=\pm 1$ to: $[{\cal C}_{k}(1)]^2 = 1 - h^2 \sin^2(k)/\varepsilon_k^2$ and $[{\cal C}_{k}(-1)]^2 = {\cal C}_{k}(1)^2 + 2 \sin{(2k)} a_k b_k/\varepsilon_k^2$. This leads to the following expressions for the widths in the diagonal ensemble
\begin{eqnarray}
\Sigma_{\hat S_j^x \hat S_{j+1}^x,{\rm DE}}^2 &=& \frac{1}{4L^2}  \sum_{k \in {\cal K}^{(+)}} 4 \alpha_k (1-\alpha_k)  \left( 1 - \frac{h^2\sin^2(k)}{\varepsilon_k^2} \right) \label{SigmaSxSx1} \\
\Sigma_{\hat S_j^y \hat S_{j+1}^y,{\rm DE}}^2 &=& \Sigma_{\hat S_j^x \hat S_{j+1}^x,{\rm DE}}^2 +  \frac{1}{4L^2}  \sum_{k \in {\cal K}^{(+)}} 4 \alpha_k (1-\alpha_k) \frac{ 2 \sin{(2k)} a_k b_k}{\varepsilon_k^2}. \label{SigmaSySy1}
\end{eqnarray}
\begin{figure}[!t]
\includegraphics[width=0.99\textwidth]{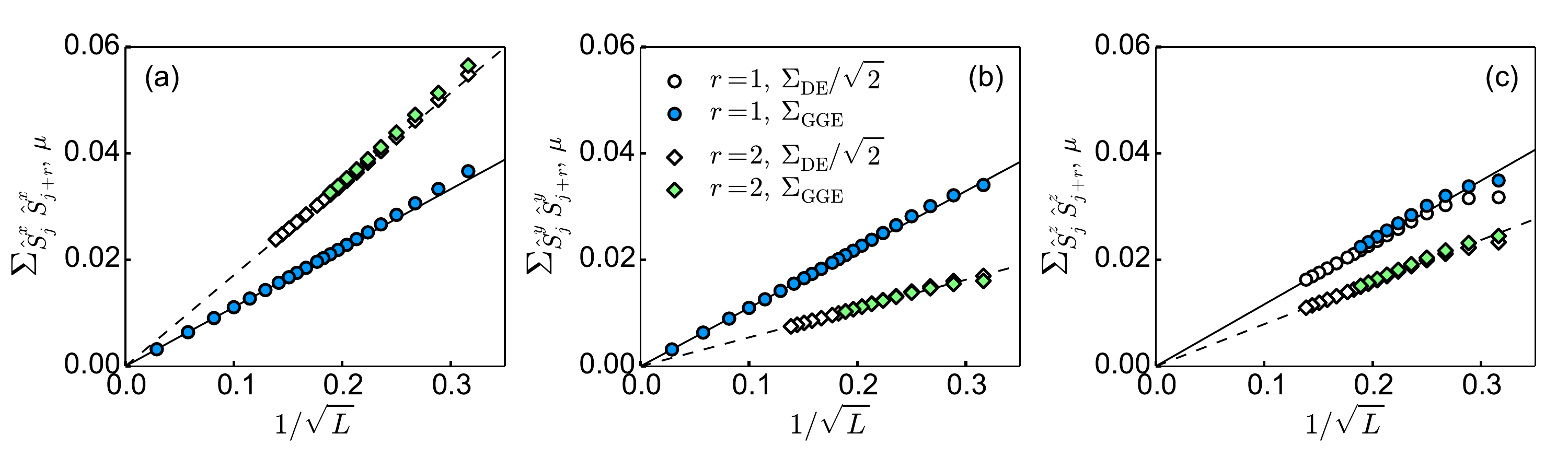}
\caption{
{\it Width $\Sigma_{\hat{\cal O}, \mu}$ in the diagonal ensemble and in the GGE vs $1/\sqrt{L}$.}
We quench from the ground state for $h_0=4.0$ to $h=0.8$. For the diagonal ensemble, we plot $\Sigma_{\hat{\cal O}, {\rm DE}}/\sqrt{2}$.
The straight lines are functions $\gamma_{a,r}/\sqrt{L}$ ($a =x,y,z$).
For $a=x,y$ and $r=1$ (solid lines), the parameter $\gamma_{a,r}$ is given by expressions in Eqs.~(\ref{SigmaSxSx1b}) and~(\ref{SigmaSySy1b}), yielding
$\gamma_{x,1}  = 0.111$  and $\gamma_{y,1}  = 0.110$.
For all the other curves, $\gamma_{a,r}$ is obtained by fitting $\Sigma_{\hat{\cal O}, {\rm DE}}/\sqrt{2}$ for $L \geq 30$.
For $r=1$ (solid line), we get $\gamma_{z,1}  = 0.116$ in panel (c). 
For $r=2$ (dashed lines), we get $\gamma_{x,2}  = 0.171$ in panel (a), $\gamma_{y,2}  = 0.054$ in panel (b), and $\gamma_{z,2}  = 0.079$ in panel (c).
}
\label{fig_width_SaSa_080}
\end{figure}
\begin{figure}[!t]
\includegraphics[width=0.99\textwidth]{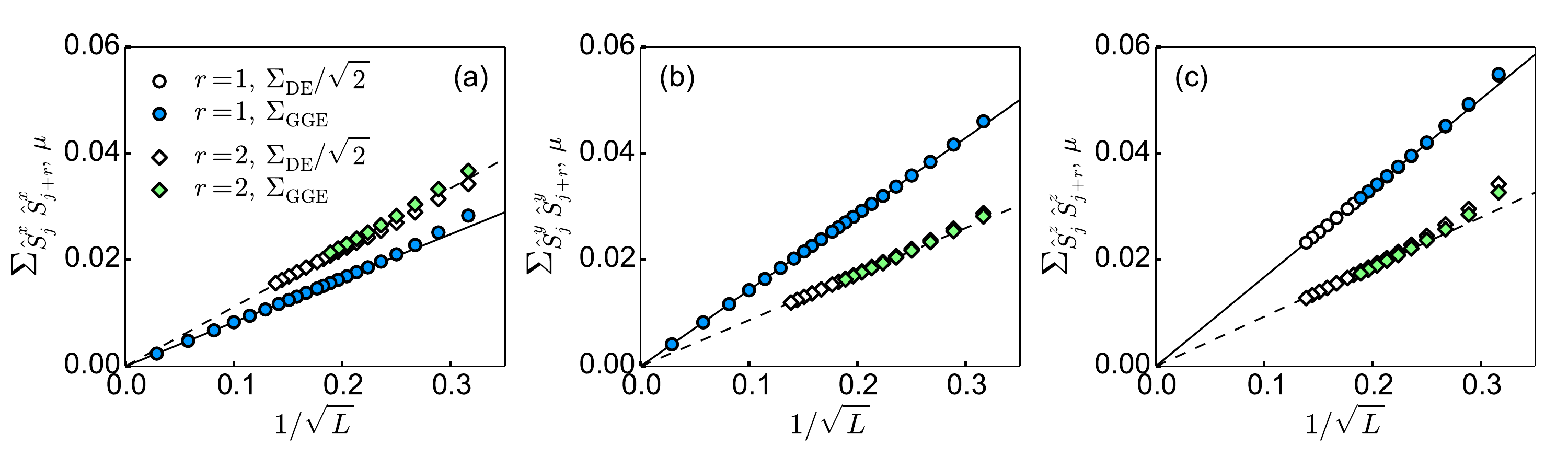}
\caption{
{\it Width $\Sigma_{\hat{\cal O}, \mu}$ in the diagonal ensemble and in the GGE vs $1/\sqrt{L}$.}
We quench from the ground state for $h_0=0.1$ to $h=1.5$. For the diagonal ensemble, we plot $\Sigma_{\hat{\cal O}, {\rm DE}}/\sqrt{2}$.
The straight lines are functions $\gamma_{a,r}/\sqrt{L}$ ($a =x,y,z$).
For $a=x,y$ and $r=1$ (solid lines), the parameter $\gamma_{a,r}$ is given by expressions in Eqs.~(\ref{SigmaSxSx1b}) and~(\ref{SigmaSySy1b}), yielding
$\gamma_{x,1}  = 0.083$  and $\gamma_{y,1}  = 0.143$.
For all the other curves, $\gamma_{a,r}$ is obtained by fitting $\Sigma_{\hat{\cal O}, {\rm DE}}/\sqrt{2}$ for $L \geq 30$.
For $r=1$ (solid line), we get $\gamma_{z,1}  = 0.167$ in panel (c). For $r=2$ (dashed lines), we get $\gamma_{x,2}  = 0.112$ in panel (a), $\gamma_{y,2}  = 0.087$ in panel (b), and $\gamma_{z,2}  = 0.093$ in panel (c).
}
\label{fig_width_SaSa_150}
\end{figure}
As for the transverse magnetization, the ratio between the widths in the diagonal ensemble and in the GGE equal
\begin{equation} \label{SigmaO_de_gge}
\frac{\Sigma_{\hat {\cal O},{\rm DE}}}{\Sigma_{\hat {\cal O},{\rm GGE}}} = \sqrt{2},
\end{equation}
for $\hat {\cal O} = \hat S_j^x \hat S_{j+1}^x$ and $\hat {\cal O} = \hat S_j^y \hat S_{j+1}^y$. This result is independent of the system size. Equations~(\ref{SigmaSxSx1}) and~(\ref{SigmaSySy1}) can also be evaluated analytically in the continuum limit
\begin{equation} \label{SigmaSxSx1b}
\hspace*{-2.2cm}
\Sigma_{\hat S_j^x \hat S_{j+1}^x,{\rm DE}}^2 =
\left\{
\begin{array}{lll}
\frac{1}{64L} \left[ 1+ h_0^2 -  \frac{2 h_0}{h} \right] & {\rm if} & h>1, \; h_0 < 1 \\
\frac{1}{64L} \left[ 4 - 3h^2 - \left( \frac{h}{h_0} \right) (4 - 2h^2) + \left( \frac{h}{h_0} \right)^2 \right] & {\rm if} & h<1, \; h_0 > 1
\end{array} \right. ,
\end{equation}
and
\begin{equation}  \label{SigmaSySy1b}
\hspace*{-2.5cm}
\Delta \Sigma_{\hat S_j^y \hat S_{j+1}^y,{\rm DE}}^2 =
\left\{
\begin{array}{lll}
\frac{1}{64L}  \left[ \frac{2h_0(2-h_0^2)}{h} + \frac{6-5h_0^2 + h_0^4}{h^2} - \frac{2h_0(4-h_0^2)}{h^3} - \frac{5-3h_0^2}{h^4} + \frac{4 h_0}{h^5} \right]  & {\rm if} & h>1, \; h_0 < 1 \\
\frac{1}{64L} \left[ 3h^2 - 2 +  \frac{2h (1-h^2) }{h_0} - \frac{1+h^2}{h_0^2}+ \frac{1}{h_0^4} \right] & {\rm if} & h<1, \; h_0 > 1
\end{array} \right. ,
\end{equation}
where $\Delta \Sigma_{\hat S_j^y \hat S_{j+1}^y,{\rm DE}}^2 =  \Sigma_{\hat S_j^y \hat S_{j+1}^y,{\rm DE}}^2 - \Sigma_{\hat S_j^x \hat S_{j+1}^x,{\rm DE}}^2$. In all cases, the widths vanish with increasing system size as
$\Sigma_{\hat {\cal O},{\rm DE}} = c/\sqrt{L}$. For the spin correlations in the $x$ direction, the largest prefactor is $c=1/4$, obtained when quenching from the initial $h_0 > 1$ to the final field $h\to 0$. For the spin correlations in the $y$ direction, the largest prefactor is $c=(1/8)\sqrt{14/5}\sim 0.209$, obtained when quenching from $h_0 \sim 0$ to $h = \sqrt{5/3}$.

In Figs.~\ref{fig_width_SaSa_080} and~\ref{fig_width_SaSa_150}, we report results for six different observables $\hat S_j^a \hat S_{j+r}^a$, $a=\{ x,y,z \}$ and $r=\{ 1,2 \}$, and two different quenches. The results for $\hat S_j^x \hat S_{j+1}^x$ and $\hat S_j^y \hat S_{j+1}^y$ were obtained using Eqs.~(\ref{SigmaSxSx1}) and (\ref{SigmaSySy1}), while the results for the other correlations were obtained through a brute force numerically evaluation of the diagonal ensemble and the GGE. We rescaled the data for the diagonal ensemble as $\Sigma_{\hat {\cal O},{\rm DE}} /\sqrt{2}$ to show the data collapse (when present). Interestingly, the expectation values obtained for $\hat S_j^z \hat S_{j+1}^z$, and all the next-nearest-neighbor spin-spin correlations, reveal that Eq.~(\ref{SigmaO_de_gge}) is not exactly fulfilled for all correlations in finite systems. Nevertheless, we find that $\Sigma_{\hat {\cal O},{\rm DE}} / \Sigma_{\hat {\cal O},{\rm GGE}} \rightarrow \sqrt{2}$ with increasing system size. 

In all cases studied in Figs.~\ref{fig_width_SaSa_080} and~\ref{fig_width_SaSa_150}, the width of the  distribution of the observables vanishes with increasing system size a $1/\sqrt{L}$. Since the width of the energy distribution also vanishes as $1/\sqrt{L}$, this implies that the states that determine the outcome of the predictions of the diagonal ensemble and the GGE in the thermodynamic limit are located in the same point in the plane defined by the eigenstate expectation values of the observable and the energies. The exact distribution of weights in each ensemble plays no role. This is the reason why, in general, the GGE can predict the expectation values of observables after relaxation despite the fact that the number of parameters one specifies for the GGE is exponentially smaller than the number of parameters one specifies for the diagonal ensemble.

\section{Summary and discussion} \label{conclusion}

In summary, we have reviewed evidence that the GGE is the appropriate statistical ensemble to describe few-body observables after quantum quenches in different families of integrable models. We focused on two models and in finite systems, for which it can be shown that the GGE describes stationary values of few-particle (spin) correlations in the entire system: the XX model and the transverse field Ising model. For the XX model, we discussed several instances in which efficient numerical calculations allow one to demonstrate that generalized thermalization occurs in quenches in the absence of translational symmetry. For the transverse field Ising model, we focused on the comparison between the expectation values in the diagonal ensemble and the GGE after quenching the transverse field in translationally invariant systems. For several observables, we proved analytically that the difference between the two ensembles vanishes in the thermodynamic limit.

Even though these models can be mapped onto noninteracting spinless fermions, we argued that there is a fundamental difference between noninteracting fermions and the aforementioned models with respect to generalized thermalization. For noninteracting fermions, the time-averaged values of all one-body observables after a quench is given by the GGE, but some extensive sets of those observables may not equilibrate. In the interacting models, the average values of all one-body observables is also given by the GGE, and we argued that there are no extensive sets of one-body observables that do not equilibrate (in the absence of real space localization due to disorder or quasi-periodic potentials \cite{gramsch12}).

The conserved quantities we utilized to construct the GGE were the occupations of single-particle eigenstates in the noninteracting fermionic models to which spins (hard-core bosons) can be mapped. As mentioned before, the onset of generalized thermalization did not depend on whether initial state or the final Hamiltonian exhibited translational invariance. This observation applies to systems in the presence of quasi-periodic potentials, but provided the potential strength is below the critical value needed for localization~\cite{gramsch12}. In contrast, as shown in Ref.~\cite{gramsch12}, the GGE fails to predict the expectation values of observables after relaxation in the localized phase.

\begin{figure}[!t]
\includegraphics[width=1\textwidth]{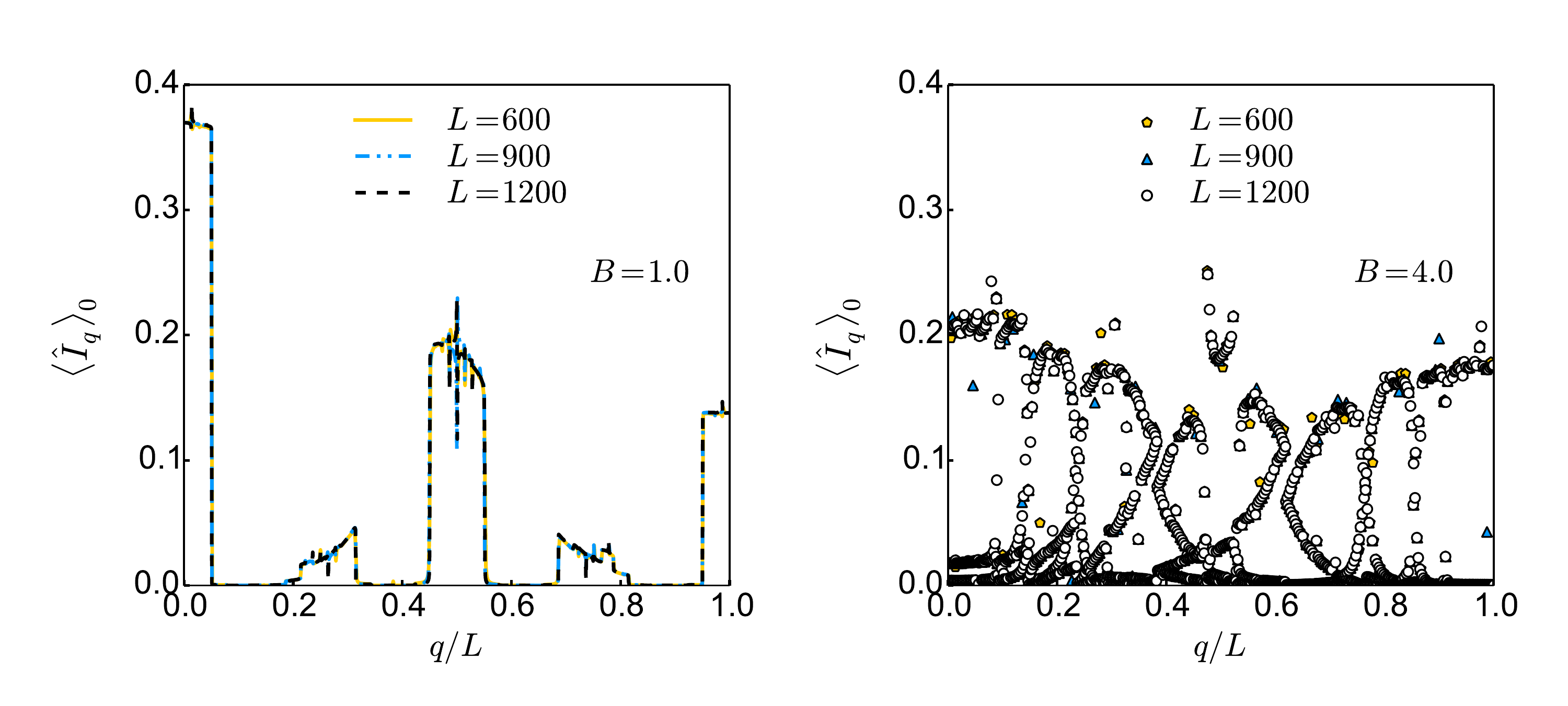}
\caption{
{\it Distribution of conserved quantities in the XX model after a quantum quench that introduces a quasi-periodic potential.}
The conserved quantities $\langle \hat I_q \rangle_0$ in the XX model are the occupations of the single-particle eigenstates of the fermionic Hamiltonian, where $q=1,2,\ldots,L$. The curves display $\langle \hat I_q \rangle_0$ after a quench from the ground state of $H_{\rm XX}$~(\ref{Hxx_def}) in the presence of a (superlattice) potential $V_j = A \tilde J \cos(\frac{2\pi j}{T})$ with period $T=4$ and amplitude $A=8$. The quench consists of turning off the superlattice potential and turning on a quasi-periodic potential $V_j = B \tilde J \cos(2\pi \sigma j)$, with $B=1$ (left panel) and $B=4$ (right panel). Quasi-periodicity (sometimes referred to as quasi-disorder) is achieved by taking $\sigma = (\sqrt{5}-1)/2$. The transition between the delocalized and the localized phase occurs for $B_c=2$. The average site occupancy is $N/L= 1/20$. The $q$-values are ordered with increasing eigenenergies of the single-particle eigenstates.
}
\label{fig_Ikdisorder}
\end{figure}

The failure of the GGE in the presence of localization can be attributed to the breakdown of statistical independence of macroscopic subsystems in the GGE, which results in the breakdown of the GGE description. This is due to the fact that conserved quantities in the localized phase are local but nonextensive, and cannot be thought of as extensive in a coarse grained way. In Fig.~\ref{fig_Ikdisorder}, we show conserved quantities in the XX model (ordered according to increasing eigenenergies), for the same initial state as in the left panel of Fig.~\ref{fig_Ik}, but now the quench consists of turning off the initial superlattice potential and turning on a quasi-periodic one. Results are shown for two strengths of the quasi-periodic potential after the quench. In the left panel in Fig.~\ref{fig_Ikdisorder}, the system is delocalized after the quench, while in the right panel in Fig.~\ref{fig_Ikdisorder} the system is localized. Note that in the former case the results are qualitatively similar to those in the left panel of Fig.~\ref{fig_Ik}, while in the latter case they are starkly different. Coarse graining in the presence of localization leads to loss of information and cannot be carried out to generate effectively extensive quantities that are meaningful. Similarly, the GGE description is not expected to apply to many-body localized systems, in which the conserved quantities are also local and nonextensive \cite{nandkishore_huse_review_15, altman_vosk_review_15}. 

Another topic that we discussed in this review is the microscopic origin of generalized thermalization in integrable systems. Our results for the transverse field Ising model, and previous studies on the XX model, show that the GGE and the diagonal ensemble predict identical expectation values of observables in the thermodynamic limit despite the fact that they are constructed using vastly different numbers of constraints. A spectral decomposition reveals that most of the weight in the diagonal ensemble and the GGE is carried by states that have similar expectation values of few-body (spin) observables. Since those states also have similar distributions of conserved quantities, these results support the hypothesis that generalized eigenstate thermalization is a generic feature in integrable systems, and calls for equivalent studies in integrable models that cannot be mapped onto noninteracting ones. Generalized eigenstate thermalization provides a microscopic understanding for the success of the GGE when traditional ensembles of statistical mechanics fail.

\section*{Acknowledgments}

This work was supported by the US Office of Naval Research. The authors are grateful to Michael Brockmann, Jean-S\'ebastien Caux, Miguel A. Cazalilla, Jacopo De Nardis, Vanja Dunjko, Fabian H. L. Essler, Maurizio Fagotti, Fabian Heidrich-Meisner, Marcin Mierzejewski, Alejandro Muramatsu, Giuseppe Mussardo, Maxim Olshanii, Anatoli Polkovnikov, Lea F. Santos, Mark Srednicki, and Tod M. Wright for many discussions and collaborations on the topics of this review.

\section*{References}

\bibliographystyle{biblev1}
\bibliography{references}

\end{document}